\documentclass[11pt]{article}

\usepackage[preprint]{acl}

\usepackage{times}
\usepackage{latexsym}

\usepackage[T1]{fontenc}
\usepackage[utf8]{inputenc}

\usepackage{microtype}

\usepackage{inconsolata}

\usepackage{amsmath}
\usepackage{graphicx}

\usepackage{hyperref}
\usepackage{url}
\usepackage{cleveref}
\usepackage{amssymb}
\usepackage{booktabs}
\usepackage{enumitem}
\usepackage{xspace}
\usepackage{multicol}
\usepackage{multirow}
\usepackage{graphicx}
\usepackage{bbm}

\usepackage{colortbl}
\definecolor{Gray}{gray}{0.9}

\newcommand{\benchmark}{DecArena\xspace}
\newcommand{\method}{GraphHarness\xspace}

\title{\benchmark: Evaluating Large-Scale Decentralized Multi-Agent Collaboration}

\author{
 \textbf{Jiali Cheng\textsuperscript{1}\thanks{Work done during internship at Amazon.}},
 \textbf{Oleg Rokhlenko\textsuperscript{2}},
 \textbf{Joo-Kyung Kim\textsuperscript{2}},
 \textbf{Hadi Amiri\textsuperscript{1}},
 \textbf{Xingzhi Guo\textsuperscript{2}}
\\
\\
 \textsuperscript{1}University of Massachusetts Lowell,
 \textsuperscript{2}Amazon Alexa AI
\\
 \small{
   \texttt{\{jiali\_cheng,hadi\_amiri\}@domain, \{olegro,jookyk,guxingzh\}@amazon.com}
 }
}

\begin{document}
\maketitle

\begin{abstract}
% \jacky{Add a line to emphasize that multi-agent systems are emerging now that personal assistants are being democratized.}
Emerging personal AI agents raise fundamental scalability and privacy concerns for centralized multi-agent systems (MASs), which rely on a global orchestrator with broad access to agents, resources, and communication. 
In decentralized environments, global visibility is unavailable. Agents need to proactively discover non-local resources (e.g. tools, skills), and collaborate with other agents to solve tasks. 
We study decentralized multi-agent collaboration under graph-constrained communication, a natural reflection of human structures. Each agent observes only its local state and can interact only with neighboring agents.
% 
% We study the problem of decentralized multi-agent collaboration, where agents have only local observations and communication is restricted to graph neighbors. 
% \jacky{add why graph - natural extension of social nets/citation nets, etc, agents can leverage the existing structure produced from human activity as a warm-up start, and add current challenges - e.g., no dataset} 
% 
% We formulate this setting as a Graph-Constrained Decentralized Partially Observable Markov Decision Process (G-Dec-POMDP) and 
We introduce a benchmark with synthetic scalable distributed tool calling with 100K+ tools. 
Our experiments reveal that recent LLM agents struggle to 
% proactively\jk{why should  be "proactively"?} 
discover non-local relevant resources and collaborate in decentralized setting.
% \jacky{is it an observation?}. 
Moreover, standard reinforcement learning fails to reliably overcome this limitation, 
% \jacky{did we demostrate it?}
% suggesting that decentralized resource discovery requires more explicit mechanisms for structured collaboration. \jacky{how about -  "suggesting that decentralized MAS needs more explicit harness mechanisms fo effective collaboration, even using SOTA LLMs"?}
suggesting that decentralized MAS needs more explicit harness mechanisms for effective collaboration, even using SOTA LLMs.
We then present a strong baseline \textbf{\method}, combining topology-aware warmup with self-improving LLM-based routing. 
% Unlike broadcasting, \method seeks relevant resources through selective local communication and learns from the success and efficiency of previous routes. 
Results show that \method substantially improves task success (+48.5) and produces higher quality collaboration (+30.6), while using fewer costs (+34.0) than baselines. 
% \jacky{cite results here}

\end{abstract}
\section{Introduction}
\label{sec:intro}
% \jk{Put some citations in Intro section.}

% \jacky{Add a few sentences to situate the context : 1. personal AI agent is being democratized rapidly (e.g., Meta Muse, OpenClaw) 2. similar to human society, it creates a new opportunity for AI agents to collabrate, coordinate on behalf of human over the existing human structural organization (e.g., social network, academic citation network, etc)}

Personal AI agents are rapidly becoming accessible to a broad population, such as Meta Muse\footnote{https://about.fb.com/news/2026/09/introducing-muse-personal-ai-agent/} and OpenClaw.\footnote{https://openclaw.ai/} The proliferation of these independently operated agents creates a new opportunity for large-scale decentralized multi-agent collaboration~\citep{kim2025towards,mao2025agentkernel}, through existing organizational structures such as social networks~\citep{jiang2026humans}.

In such multi-agent systems (MASs), agents are naturally organized in a decentralized manner, without a centralized orchestrator~\citep{li2023camel,liu2024autonomous}. Individual agents may differ in their knowledge, capabilities, and access to resources, making collaboration necessary for solving tasks that no single agent can complete alone~\citep{mao2026decentralized,cheng2025webatlas}. Agents need to discover relevant peers, exchange information, and coordinate actions autonomously.
% External resources such as reusable tools and skills further increase this heterogeneity, providing additional capabilities that may remain distributed across different agents.

% Large language model (LLM) agents increasingly extend their capabilities through \emph{external knowledge}, such as external documents, tools, and skills.
% modular, named, and reusable procedures that can be invoked at inference time. 
% A knowledge may encode a workflow, operate an external tool, query a specialized data source, or provide domain-specific expertise. Emerging systems already contain tens of thousands of such capabilities, in centralized format, knowledge become a standard unit of capability sharing across users, organizations, and agent platforms.\jk{Not sure about this sentence. Rephrase it.}

% Existing systems typically assume that knowledge are organized through a \emph{centralized registry}. At task time, an agent or an external retriever searches a globally visible catalog and selects one or more candidate knowledge based on their descriptions~\cite{toolformer,react,toolllm,knowledgerouter2026}. This abstraction has enabled substantial progress in tool selection and knowledge routing. However, it relies on several strong assumptions: the registry is complete, every capability is globally visible, knowledge descriptions are centrally indexed, and a retriever can identify the relevant capability before the agent begins acting.

Prevailing MASs are mostly centralized, with three major limitations.
% of the prevailing centralized paradigm. 
% \jacky{ change the framing: we argue that the centralized MAS will be bottleneck by 1)skill maintainability 2)privacy concern 3)   }
\emph{1) Difficult to scale and maintain.}
% \jk{Don't break intro section with paragraphs like this. It's better just to list in plain texts. If you want separate emphaisis, do it in other sections.}
Existing MASs assume that knowledge is stored in a centralized registry. As multi-agent ecosystems grow toward web scale, it becomes prohibitively expensive for all individual agents to access the full capability space, as resources are continuously added, updated, deprecated, or moved across organizations~\citep{grotschla2025agentsnet,zhu2025multiagentbench}. 
% As a result, centralized registries struggle to represent dynamic capability ecosystems in which resources evolve over time and remain distributed across independently managed agents.
% 
\emph{2) Privacy concern.}
% A global registry also conflicts with privacy, ownership, and execution constraints. 
Many useful skills depend on private data, restricted information, or specialized runtime, which cannot always be copied into a shared repository. 
Even when a resource can be advertised, its implementation or intermediate states may need to remain hidden from other parties~\citep{louck2025improving,zou2025blocka2a,hu2025inter,cheng2025tool}. 
% In these settings, an agent may be willing to reveal a resource description, return an execution result, or forward a request, but not disclose the underlying resource itself. This motivates a decentralized setting in which capabilities remain locally owned and are accessed through controlled interaction.
% 
\emph{3) Less collaborative and proactive in decentralized environment.}
Current MASs typically rely on centralized orchestrators to organize agents and select relevant resources. 
% This reduces resource use to a one-step retrieval problem: the agent receives a task, queries an index, and invokes the selected item. 
However, when no orchestrator exists, individual agents must proactively coordinate with other agents and discover useful resources through interaction~\citep{ma2025hunger,davidson2026collaboration,pappu2026multi}. 
% It must decide which neighbors to query, whether to forward requests, and how to evaluate returned descriptions or outputs. 
% Resource use therefore becomes a problem of proactive discovery, routing, and decentralized collaboration rather than passive retrieval from a centralized catalog.

We study the following research question:

\begin{quote}
\emph{Can agents proactively discover external knowledge and solve complex tasks in a large-scale decentralized manner?}
\end{quote}

\begin{figure*}
    \centering
    \vspace{-25pt}
    \includegraphics[width=0.98\linewidth]{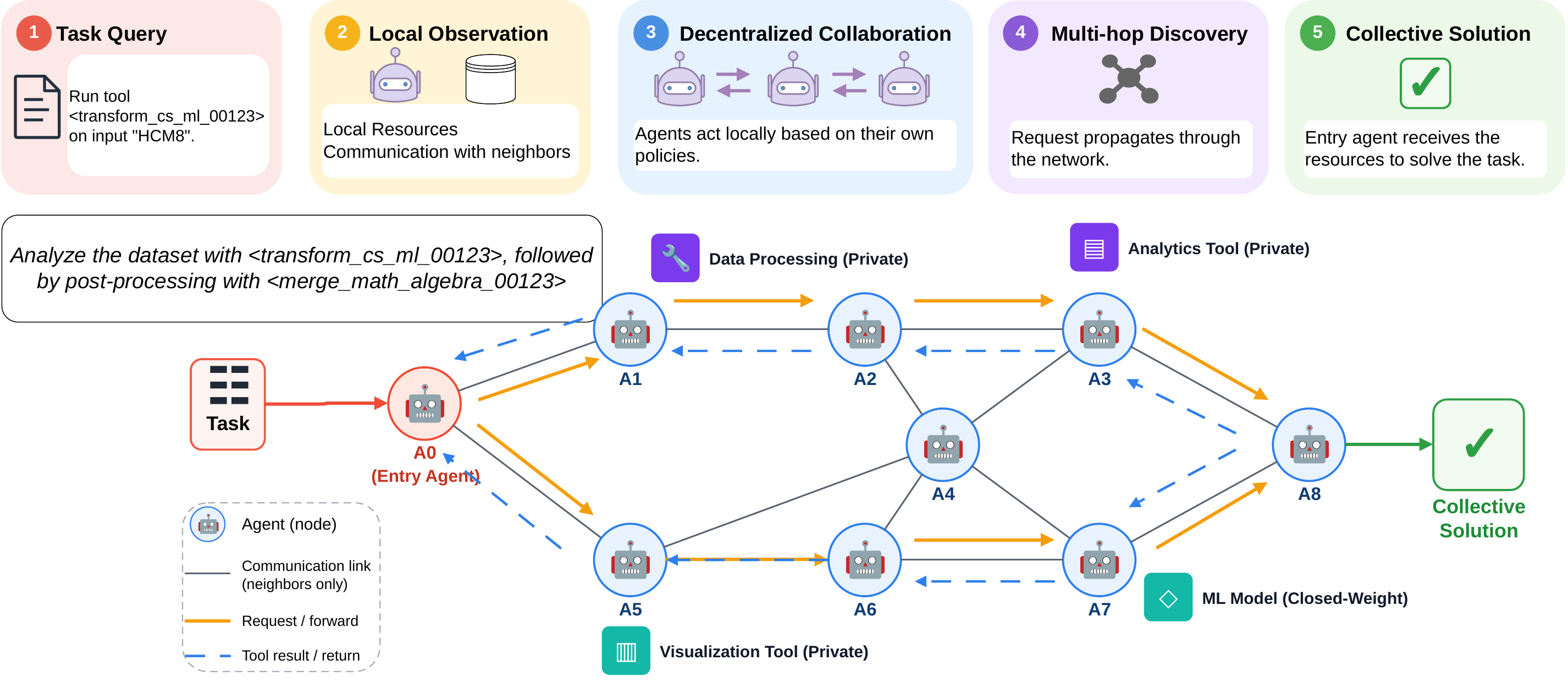}
    \caption{\textbf{Illustration of decentralized multi-agent collaboration (Graph-Constrained Dec-POMDP) in DecArena.} An entry agent receives a task and can access only its own local resources and immediate neighbors. Each agent decides locally whether to execute an owned resource, return a result, reject the request, or forward it to neighboring agents. Through multi-hop neighbor-to-neighbor communication, the request propagates across the network until the required distributed resources are discovered and composed, after which the final result is returned to the entry agent as the collective solution.}
    \label{fig:fig1}
\end{figure*}

% \paragraph{Our setting: decentralized multi-agent collaboration.}
We study \textbf{decentralized multi-agent collaboration}, a novel regime to evaluate MAS on decentralized knowledge discovery and collaboration. 
% In this setting, agents are connected through a communication graph. 
% Each agent possesses only a small set of locally visible knowledge. 
In this setting, each agent only observes its local knowledge and its immediate neighbors on the communication graph, with zero global knowledge.
% It does not have access to a global index of knowledge. 
Given a query, the originating agent may choose to proactively search through the network by querying neighbors, forwarding requests, and deciding whether to terminate or continue the search.
% Queries originate at agents that may not possess the required knowledge. The originating agent must locate and access the relevant capability through a sequence of local communication and routing decisions. 
A successful task execution requires involved agents to collectively and efficiently locate the required capability by the query.
% The second is \emph{structural}: the agent must determine where that capability can be reached in the communication network.

% \paragraph{Significance.}
This setting extends multi-agent collaboration from coordination of centralized orchestrator to discovery over independently operated peers. Each agent develops a partial, local, and workload-dependent understanding of the surrounding capability network through interaction. 
This extension is necessary when capabilities cross organizational boundaries, ownership changes over time, and relevant expertise may be reachable only through intermediate collaborators. 
% By providing a controlled substrate in which knowledge placement, network topology, exchange mechanism, and discovery policy can be studied independently, 

We design a benchmark for the proposed setting with 160 tasks of three difficulty levels, with 100k+ distributed tools, five network scales and three topologies. We propose three metrics 1) task success, 2) collaboration quality, 3) efficiency.

We summarize our contributions as follows.

\begin{itemize}[leftmargin=1.4em,itemsep=0.25em,topsep=0.25em]

\item \textbf{A formulation for decentralized knowledge discovery and collaboration.}
We introduce the Graph-Constrained Dec-POMDP setting where knowledge are distributed across a network of agents rather than stored in a globally searchable registry. Agents must proactively locate relevant capabilities through local interaction and access them through either knowledge transfer or remote execution.

\item \textbf{Insights on multi-agent behaviors.}
We discovery when and why agents fail in decentralized multi-agent collaboration, which sheds light on the future development of agent, including training data and evaluation tasks.

\item \textbf{A baseline agent learned proactive collaborate.}
We design a baseline agent that can learn from past experiences and evolve over time. 
% (Can we formulate this as RSI?)
% \jk{Could be somewhat related to RSI but not sure how effectively link with RSI yet.}

\end{itemize}

% AgentNet: Decentralized Evolutionary Coordination for LLM-based Multi-Agent Systems, 10, Tasks: math, function calling, QA. Does not require collaboration.

% MultiAgentBench: Evaluating the Collaboration and Competition of LLM agents, 7, Tasks: coding, db error analysis, minecraft.

\section{Related Work}
% \jk{How bout moving this to section 2 after Introduction?}

\paragraph{Multi-agent collaboration.}
LLM-based multi-agent systems (MAS) have been studied for collaborative problem solving, where agents specialize, communicate, and reach consensus~\citep{guo2024large,tran2025multi}. Prior work examines collaboration under information asymmetry~\citep{liu2024autonomous}, scientific idea generation~\citep{su2025many}, self-organizing role assignment~\citep{chen2024sagents}, and delegation among agents and humans~\citep{tomasev2026intelligent,zhu2026choose}. Yet the benefits of adding more agents are not guaranteed: several studies show that naive multi-agent teams can undermine strong individual agents~\citep{pappu2026multi}, fail to explore one another's knowledge~\citep{choi2026multi}, exhibit collaboration gaps in open-world settings~\citep{davidson2026collaboration}, and expose weak links under adversarial prompting~\citep{arora2025exposing}. This motivates mechanisms that make collaboration robust rather than merely larger.

A growing body of work optimizes \emph{how} agents are connected. Methods learn or prune communication graphs~\citep{li2025adaptive} and jointly optimize prompts and topologies~\citep{zhou2026multi}. Routing approaches decide which agent or context to invoke~\citep{liu2025rcr,zhang2025agentrouter,shi2025ngrouter,yang2026augmented}, including RL-trained routers~\citep{zhang2025routerr1} and decentralized routing policies~\citep{lozanocuadra2025learning}. Decentralized designs coordinate agents without a central controller~\citep{yang2025agentnet,mao2026decentralized}, and budget-aware formulations trade off performance against cost~\citep{yang2026bamas}. Our skill-graph formulation is closest to this line, treating topology as a first-class, adaptable object.

\paragraph{Learning and self-evolution.}
Beyond fixed pipelines, agents can be trained to collaborate via multi-agent reinforcement learning~\citep{liu2025llm,liao2025marft,feng2026drmas,liu2026spiral}, and can self-improve through shared experience~\citep{weng2026group}, scaling~\citep{qian2025scaling,shao2026monoscale}, or test-time adaptation~\citep{liu2026masonthefly}. Most relevant to us, skill-centric systems co-evolve reusable competencies across agents~\citep{pan2026skillmas,nie2026skillgraph,zhang2026swarm}.

\paragraph{Benchmarks and applications.}
MAS are increasingly evaluated on standardized benchmarks spanning general collaboration and competition~\citep{zhu2025multiagentbench}, cooperative RL tasks~\citep{papadopoulos2025extended}, planning and scheduling~\citep{geng2026realm}, enterprise workflows~\citep{yu2026beyond}, multimodal embodied settings~\citep{long2024teamcraft}, competitive games~\citep{wang2024battleagentbench}, swarm intelligence~\citep{ruan2025benchmarking}, and medical tasks~\citep{zhu2025medagentboard}; questions of dependency trust are also raised~\citep{yao2025can}. Applications range from autonomous scientific discovery~\citep{ghareeb2026multi,jin2025aleks,lyu2026evoscientist} and infrastructure management~\citep{saifullah2024multi} to large-scale social, hospital, and economic simulation~\citep{li2024agenthospital,mao2025agentkernel,qi2026economy}, often built on general MAS platforms~\citep{gao2024agentscope}.

\section{\benchmark: A Benchmark for Decentralized Multi-Agent Collaboration}
\label{sec:method}

We introduce a benchmark for evaluating large-scale decentralized collaboration among agents, where no agent has access to a global registry of resources. Instead, resources are distributed, and agents must solve tasks using only their own local resources and information obtained through communication. The benchmark therefore tests whether agents can discover, request, route to, and compose distributed resources under local observability.

\subsection{Problem Formulation}
% : Graph-Constrained Decentralized POMDP}
\label{sec:problem-formulation}

% We study decentralized multi-agent collaboration over a communication graph. A task is given to one \textbf{entry agent}. The resources required to solve the task may be owned by other agents, so the entry agent must discover and access them through neighbor-to-neighbor communication.

% \paragraph{Example.}
% Suppose agent $A_1$ receives the instruction ``apply \texttt{tool\_123} to \texttt{input\_1},'' but does not own \texttt{tool\_123}. Agent $A_1$ asks its neighbors $A_2$ and $A_3$. Agent $A_3$ rejects the request. Agent $A_2$ does not own the tool, so it forwards the request to $A_4$. Agent $A_4$ owns the tool, executes \texttt{tool\_123(input\_1)}, and returns the result through $A_2$ to $A_1$. The task finishes when $A_1$ receives the result and produces the final answer.

% This setting requires agents to discover non-local resources, route requests over multiple hops, handle selective cooperation, and solve the task using only local information.

\paragraph{Decentralized POMDP}
\label{sec:dec-pomdp}

A Decentralized Partially Observable Markov Decision Process (Dec-POMDP) models multiple agents that act under partial observability and optimize a shared objective~\citep{bernstein2002complexity,oliehoek2016concise}. We use the tuple $\left(\mathcal{A}, \mathcal{S}, \{\mathcal{U}_i\}_{i=1}^{n}, T, R, \{\mathcal{O}_i\}_{i=1}^{n}, O\right)$, where
\begin{itemize}
\itemsep0pt
    \item $\mathcal{A}=\{A_1,\ldots,A_n\}$ is the set of agents
    \item $\mathcal{S}$ is the global environment-state space
    % \item $\mathcal{U}_i$ is the action space of agent $A_i$, and $\mathcal{U}=\prod_i \mathcal{U}_i$ is the joint-action space\jk{Not sure if this is a correct expression as this may look like just multiplications.}
    \item $\mathcal{U}$ is the action space
    % of agent $A_i$, and $\mathcal{U}=\prod_i \mathcal{U}_i$ is the joint-action space\jk{Not sure if this is a correct expression as this may look like just multiplications.}
    \item $T(s'\mid s,\mathbf{u})$ is the transition model under joint action $\mathbf{u}$
    \item $R(s,\mathbf{u})\in\mathbb{R}$ is the shared reward
    \item $\mathcal{O}_i$ is the local-observation space of agent $A_i$
    \item $O$ defines the observation distribution after an environment transition
\end{itemize}

The key property is decentralization. There is no controller that observes the complete state and chooses actions for all agents. Instead, each agent acts from its own local history $h_{i,t}=\left(o_{i,1},u_{i,1},\ldots,o_{i,t-1},u_{i,t-1},o_{i,t}\right)$ using a local policy $u_{i,t}\sim\pi_i\left(\cdot\mid h_{i,t}\right)$

% A single-agent POMDP can be partially observable, but it is not decentralized because only one policy chooses actions. A system with a global orchestrator that observes all agents and routes every request is also not decentralized under our definition.

\paragraph{Graph-Constrained Dec-POMDP}
\label{sec:graph-dec-pomdp}

We extend the Dec-POMDP with a communication graph $G=(\mathcal{A},\mathcal{E})$, where $(A_i,A_j) \in \mathcal{E}$ means that $A_i$ and $A_j$ may communicate directly. The neighbor set of $A_i$ is $\mathcal{N}_i=\left\{A_j\in\mathcal{A}:(A_i,A_j) \in \mathcal{E}\right\}$. 
An agent can communicate only with its immediate neighbors. A request for a resource owned by a distant agent must be routed across multiple edges. Agents do not receive a global resource index, the full network topology, or the private states of other agents.

\paragraph{Local observation.}
At round $t$, agent $A_i$ observes only its local view $o_{i,t}=\left(x_{i,t},M_{i,t},\mathcal{K}_i,\mathcal{N}_i,\mathbf{1}[i=e]q\right)$, where $x_{i,t}$ is the local execution state, $M_{i,t}$ contains messages received from neighbors, $\mathcal{K}_i$ is the set of local resources owned by $A_i$, $e$ is the entry-agent index, and $q$ is the task instruction. Only the entry agent observes the full task at the start of an episode. Other agents learn about the task only through messages they receive.

\paragraph{Resources.}
Let $\mathcal{K}$ denote the full set of resources (tools).
% \jk{In this work, aren't only tools used?} 
Each resource has an owner. A resource remains local unless its owner explicitly shares its content or executes it and returns the result.

\paragraph{Tasks.}
A task is represented by $\mathcal{T}=(q,z,\mathcal{G}_{\mathcal{T}})$, where $q$ is the instruction, $z$ is the input, and $\mathcal{G}_{\mathcal{T}}$ is a directed acyclic graph of required resources. Each node corresponds to one resource call, and its input is either the original task input or the outputs of preceding nodes. Independent resource calls can be executed in parallel, while dependent calls wait for their required inputs.

% To solve the task, agents must discover the owners of the required resources, route requests through the communication graph, obtain the corresponding outputs, and return them to the entry agent. The task succeeds when the entry agent produces the correct final output.

\paragraph{Solution}

The solution is a collection of local policies $\boldsymbol{\pi}=\{\pi_i\}_{i=1}^{n}$. At each node visit, an agent may reject the request, return local resource content, return an execution result, or forward the request to one or more neighbors. Returning content, returning a result, and forwarding may be combined. Reject is exclusive.

A local policy acts from the agent's current observation, local interaction history, and optional persistent memory $W_i$ $u_{i,t}\sim\pi_i\left(\cdot\mid h_{i,t},W_i\right)$, where $h_{i,t}$ denotes agent $i$'s within-task interaction history up to time $t$, and $W_i$ denotes optional persistent memory accumulated across tasks.
% \jk{Explain what $h$ and $W$ are.}

% We evaluate policies by both task success and collaboration cost. A generic objective is

% \begin{equation}
% J(\boldsymbol{\pi})=\mathbb{E}_{\boldsymbol{\pi}}\left[r_{\mathrm{task}}-\lambda_m C_{\mathrm{msg}}-\lambda_h C_{\mathrm{hop}}-\lambda_c C_{\mathrm{call}}\right],
% \end{equation}

% where $r_{\mathrm{task}}$ measures task correctness and the cost terms measure communication, routing, and resource execution. This separates effective collaboration from exhaustive broadcast, which can eventually find a resource but becomes increasingly expensive as the network grows.

\subsection{Task Construction}
\label{sec:task}

We use synthetic tool calling as the primary benchmark because it isolates decentralized collaboration from general language-model knowledge. Tool execution is deterministic and exactly verifiable, the construction scales to more than 100k+ tools, and private keyed behavior prevents an LLM from producing the correct output without accessing the required tool.

\paragraph{Local resources.}
The atomic resource is a \emph{tool}. We define two tool types: a \emph{transform} maps one input string to one output string, while a \emph{merge} maps two or more input strings to one output string. To give semantics to the tool names, we use the OpenAlex~\citep{priem2022openalex} hierarchy of subjects. Each tool is associated with a broad field and a subfield, such as \texttt{transform\_cs\_ml\_00123} and \texttt{merge\_math\_algebra\_00123}. Tools that share ancestors in the taxonomy are semantically related, providing agents with useful but non-oracle signals for routing. The actual output of each tool is determined by a private secret key and cannot be inferred from its semantic name.

\paragraph{Tasks.}
Each task is a multi-step tool-composition pipeline over a single input string. Tools within the same step can execute in parallel, while later steps depend on outputs from earlier steps. Step~1 applies $k\in\{1,2,3\}$ transforms to the raw input, Step~2 optionally merges their outputs, and Step~3 optionally applies a final transform. We refer to these as level-1, level-2, and level-3 tasks, respectively. A task exposes the input, pipeline structure, and required tool names, while gold intermediate outputs and the final answer remain hidden and are used only for evaluation.

\paragraph{Synthetic construction.}
Tools and tasks are generated deterministically (Appendix~\ref{app:deterministic}), making the benchmark fully reproducible and exactly verifiable.
% \jk{Exactly how generated? Provide the details at least in Appendix.} 
Each tool receives a secret key derived from its identifier. Given its inputs, the tool canonicalizes and concatenates them, applies a keyed hash, and returns a fixed-length token. A single input realizes a transform, while multiple inputs realize a merge. Higher-level tasks extend lower-level task families, providing controlled compositional depth while reusing the same underlying tools. Details are presented in Appendix~\ref{app:task-construction}.

\paragraph{Task difficulty.}
The full tool bank contains more than 100k+ tools distributed throughout the subject hierarchy. We instantiate 140 evaluation tasks, including 60 level-1, 40 level-2, and 40 level-3 tasks.
% , which collectively require 186 distinct tools. 
The remaining tools form a large set of semantically structured distractors. Many share subjects or higher-level ancestors with the required tools but are not useful for the task. Consequently, semantic similarity provides a meaningful routing signal but does not identify the correct tool or its owner, requiring agents to perform targeted discovery over a large distributed capability space.

\subsection{Decentralized Environment}
\label{sec:environment}

\paragraph{Tool ownership and accessibility.}
Each active tool is assigned to exactly one agent. Since the secret key remains with the owner, a required tool can be executed only by its owner. The entry agent needs to locate the owner and request the corresponding result through the communication network.

We ensure that every benchmark task is solvable from every entry point. Each required tool has an owner in the environment, and the communication graph is connected. We then stratify entry agents by the graph distance to required tool owners to measure how collaboration changes as resources become more remote.

\paragraph{Local observability.}
Agents observe their own local resources and their immediate neighbors. There is no global tool registry, global graph view, or centralized router. Information about remote resources is obtained only through neighbor-to-neighbor communication. This prevents the benchmark from reducing to centralized retrieval over a flat tool collection.

\paragraph{Network topology and resource placement.}
% \jk{In addition to topology, need to explain how the synthesized tools are spreaded in the agents. Randomly? Based on the name hierarchy? or other ways? }
We evaluate three graph families. In a \emph{random graph}, connectivity is independent of agent capabilities. In a \emph{semantic graph}, agents are preferentially connected to others with similar capability profiles. In a \emph{scale-free graph}, preferential attachment creates a small number of hubs and many low-degree nodes. These topologies expose different routing challenges while preserving the same local interaction protocol. Resource distribution over the network and network statistics are presented in Appendix~\ref{app:network}.

\paragraph{Action space.}
When an agent receives a request, it chooses actions using only the request and its local view. The primitive actions are:
\begin{itemize}[leftmargin=1em]%,itemsep=0.25em,topsep=0.25em]
\itemsep0pt
    \item \texttt{Reject}: decline the request and return no result;
    \item \texttt{Return}: execute a local tool and return its result;
    \item \texttt{Forward}: send a request to one or more neighbors;
    \item \texttt{Submit}: stop and submit the final answer.
\end{itemize}
\texttt{Return} and \texttt{Forward} may be combined. This allows an agent to provide a local result while continuing to search for other required resources. Inefficient behavior is not prohibited and is instead reflected in the efficiency metrics.

\subsection{Evaluation}
\label{sec:metrics}

We evaluate each run along three complementary dimensions: task success, collaboration quality, and efficiency. All case-level scores are aggregated over tasks and reported as percentages.

\paragraph{Task success.}
Task success measures whether the agent system completes the full tool-composition pipeline and produces the exact gold output. Because tool execution is deterministic, final answers can be verified exactly. We additionally report per-tool output accuracy as a diagnostic of partial progress on failed tasks, indicating whether the system successfully discovered and used some required tools even when the final task was not completed.

\paragraph{Collaboration quality.}
Collaboration quality measures whether the system successfully identifies, obtains, and correctly uses the remote resources required by a task. Specifically, it combines three aspects: whether required remote tools are queried, whether their execution results are successfully returned, and whether the returned results are used correctly in the task pipeline. The collaboration score is the average of these three components. For tasks whose required tools are all local to the entry agent, no collaboration is necessary. We therefore report remote-only collaboration separately to avoid inflating the headline score with local cases.

\paragraph{Efficiency.}
Efficiency measures how economically the system explores the network while solving a task. It combines three normalized components. \emph{Visit efficiency} measures how many agent visits are used relative to a distance-based reference. \emph{Query focus} measures how selectively requests are forwarded and penalizes unnecessary fan-out. \emph{Round efficiency} measures the number of interaction rounds relative to an idealized reference that accounts for both tool distance and pipeline depth. Together, these metrics distinguish targeted collaboration from success obtained through broad or redundant exploration. Exact definitions are provided in Appendix~\ref{app:metrics}.

% Because failed runs may terminate early and appear artificially inexpensive, we report efficiency both over all runs and conditioned on successful completion. We also report raw token usage, wall-clock time, repeated visits, and rejection-related failure diagnostics.

\section{Experiments}\label{sec:exp}

We evaluate decentralized collaboration by varying where a task enters the network and how much routing state an agent can use. Unless otherwise noted, all methods operate under the same local-observation and neighbor-communication constraints defined in Section~\ref{sec:method}.

\subsection{Evaluation Protocol}
\label{sec:experimental-protocol}

\paragraph{Entry agents.}
For each task, we evaluate every agent as the entry agent. Let $\mathcal{T}(q)$ be the tools required by task $q$, let $o(t)$ be the owner of tool $t$, and let $d_{\mathcal{G}}(\cdot,\cdot)$ denote shortest-path distance in the communication graph. We define the case distance as:\looseness-1
\begin{equation}
d(q,A_0)=\max_{t\in\mathcal{T}(q)} d_{\mathcal{G}}\left(A_0,o(t)\right).
\end{equation}

This measures the distance to the farthest required tool owner and naturally separates local, one-hop, two-hop, and longer-range cases. Reporting performance by distance allows us to distinguish local tool use from increasingly difficult multi-hop collaboration.

\paragraph{Coordination protocol.}
The main benchmark uses recursive routing. When an agent receives a request, it may answer using an owned tool, reject the request, or forward it to selected neighbors. Each receiving agent makes its own local routing decision. The entry agent therefore does not observe or control every intermediate hop.

\subsection{Evaluated Agents}
\label{sec:agents}

We evaluate the following agents:
1) \textbf{Reflex Agent}~\citep{reflex}, which chooses its next action using only the current local observation, $u_{i,t}\sim\pi_{\mathrm{LLM}}(\cdot\mid o_{i,t})$; 
2) \textbf{Reflex + History}~\citep{yao2023react}, which conditions on the agent's within-task interaction history, $u_{i,t}\sim\pi_{\mathrm{LLM}}(\cdot\mid h_{i,t})$, including previously contacted neighbors and returned results; 
3) \textbf{Experience-Conditioned Agent}, which additionally retrieves successful local experiences from previous tasks, $u_{i,t}\sim\pi_{\mathrm{LLM}}(\cdot\mid h_{i,t},\mathcal{H}_i^{+})$; 
4) \textbf{Numerical Q-Routing}, a Q-Routing-based baseline~\citep{boyan1993packet} in which each agent maintains local scores for observed resource-neighbor pairs and updates them from routing outcomes; 
5) \textbf{Broadcast}, which forwards every resource request to all immediate neighbors while still allowing locally available results to be returned; 
6) \textbf{Oracle Route}, which gives only the entry agent the owner and a shortest route to each required tool; 
7) \textbf{Oracle Route All}, which exposes the same oracle routing information to every participating agent; 
8) \textbf{Collaborative Prompting}, which augments the Reflex prompt with explicit instructions to actively seek help, return useful local results, and forward requests when appropriate; 
9) \textbf{Supervised Fine-Tuning (SFT)}, which fine-tunes a shared collaboration policy on high-reward trajectories collected from the training networks using local observations and accepted structured decisions; 
10) \textbf{Group Relative Policy Optimization (GRPO)}~\citep{shao2024deepseekmath}, which starts from the SFT policy, samples multiple complete collaboration trajectories per task, and optimizes the policy using rewards based on correctness, collaboration quality, and efficiency; and 
11) \textbf{On-Policy Self-Distillation (OPSD)}~\citep{zhao2026self}, which distills a self-teacher with oracle routing information into a student policy.

Full details are provided in Appendix~\ref{app:agents}.

\section{A Strong Baseline: \method}
\label{sec:baseline}

\method is a collaboration harness that combines a topology-aware warmup with experience-adaptive LLM routing. It gives each agent useful local routing context while preserving decentralized execution. No agent receives a global resource registry or global graph view.

\subsection{Topology-Aware Warmup}
\label{sec:warmup}

Before evaluation, agents perform local message passing on auxiliary tasks that are disjoint from the test tasks. Each agent begins with a routing state $W_i^{(0)}$ derived from its own resource metadata. At warmup layer $\ell$, it exchanges compact routing information only with immediate neighbors and updates its state as
\begin{equation}
\small
W_i^{(\ell+1)}=\Phi\left(W_i^{(\ell)},\operatorname{AGG}\left\{\Psi\left(W_i^{(\ell)},W_j^{(\ell)}\right):A_j\in\mathcal{N}_i\right\}\right).
\end{equation}

After $L$ rounds, $W_i=W_i^{(L)}$ is fixed and used as persistent routing context. Information can propagate beyond one hop across warmup rounds, but every communication step still follows an edge in the graph. 
% \jk{Don't miss the correct citation.}

\subsection{Experience-Adaptive Routing}
\label{sec:routing}

During evaluation, each agent combines its warmup state with its current task history and local experience memory. Let $M_{i,t}$ denote the routing experience stored by agent $A_i$. For a request involving resource $k$ and candidate neighbor $A_j$, the LLM produces a routing score
\begin{equation}
s_i(k,A_j,t)=f_{\mathrm{LLM}}\left(k,A_j,W_i,M_{i,t},h_{i,t}\right).
\end{equation}

The agent uses these scores to choose one or more neighbors. After observing the outcome, it updates its local experience memory
\begin{equation}
M_{i,t+1}=\operatorname{Update}_{\mathrm{LLM}}\left(M_{i,t},k,A_j,r_i,o_{i,t+1}\right).
\end{equation}

The warmup provides an initial topology-aware routing prior, while the experience memory adapts from successful and unsuccessful interactions. Adaptation occurs through local context and memory rather than a centralized routing table. Additional protocol and ablation details are given in Appendix~\ref{sec:additional-experiments}.

\begin{table*}[t]
\centering
\small
\caption{Comparison of decentralized agent designs. 
% \jacky{qwen? or sonnet}\jk{Specify the LLM for each title row as it's currently confusing.} 
Detailed results with more base LLMs are shown in Appendix~\ref{app:results} Table~\ref{tab:main_full}.}
\label{tab:main}
\begin{tabular}{lcccc}
\toprule
\textbf{Agent} & \textbf{Base LLM} & \textbf{Success} & \textbf{Collaboration} & \textbf{Efficiency} \\
\midrule
\rowcolor{Gray} Global Registry  & NA & \multicolumn{3}{c}{Out of Context} \\
% \rowcolor{Gray} Global Registry + RAG & 82.6 & N/A & N/A \\
\midrule
\multicolumn{5}{c}{\textit{Static Closed-Source Agents}} \\
\midrule
% Reflex & Claude Sonnet 4.6 & 6.4 & 34.4 & 59.9 \\
% Reflex + History & Claude Sonnet 4.6 & 6.2 & 28.6 & 72.3 \\
% Broadcast & Claude Sonnet 4.6 & 71.3 & 75.3 & 36.4 \\
% Oracle Route & Claude Sonnet 4.6 & 33.5 & 62.1 & 69.5\\
% Oracle Route All & Claude Sonnet 4.6 & 27.7 & 59.3 & 67.2 \\
Reflex & QWen3.8 27B        & 5.2 & 27.5 & 37.5 \\
Reflex + History & QWen3.8 27B        & 5.0 & 30.3 & 36.1 \\
Broadcast & QWen3.8 27B & 40.6 & 68.7 & 45.7 \\
Oracle Route & QWen3.8 27B & 33.6 & 63.9 & 68.7 \\
Collaborative Prompting & QWen3.8 27B & 7.8 & 37.6 & 60.3 \\
\midrule
\multicolumn{5}{c}{\textit{Trained Open-Source Agents}} \\
\midrule
SFT  & QWen 3.5 4B & 0.8 & 48.0 & 36.3 \\
GRPO & QWen 3.5 4B & 0.5 & 46.8 & 56.6 \\
OPSD & QWen 3.5 4B & 5.7 & 46.9 & 56.4 \\
\midrule
\multicolumn{5}{c}{\textit{Self-Improving Closed-Source Agents}} \\
\midrule
% Numeric Q-Routing & Claude Sonnet 4.6 & 3.9 & 21.2 & 80.8 \\
% Success-Trajectory Memory & Claude Sonnet 4.6 & 1.1 & 31.0 & 82.3 \\
% Location Learning & Claude Sonnet 4.6 & 3.8 & 46.7 & 53.3 \\
% LLM Q-Learning & Claude Sonnet 4.6 & 19.1 & 35.9 & 63.1 \\
Numeric Q-Routing         & QWen3.8 27B & 5.0 & 22.0 & 80.1 \\
Success-Trajectory Memory & QWen3.8 27B &  1.1 & 22.3 & 62.5 \\
Location Learning         & QWen3.8 27B & 3.6 & 36.1 & 56.8 \\
\midrule
Warmup (Ours)             & QWen3.8 27B & 33.8 & 37.2 & 58.2 \\
LLM Q-Learning (Ours)     & QWen3.8 27B & 20.6 & 37.7 & 56.8 \\
% \textbf{\method} (Ours) & Claude Sonnet 4.6 & \textbf{76.8} & \textbf{78.3} & \textbf{83.8} \\
\textbf{\method (Ours)}   & QWen3.8 27B       & \textbf{53.7} & \textbf{58.1} & \textbf{71.5} \\
\bottomrule
\end{tabular}
\end{table*}

\subsection{Results}

The results in Table~\ref{tab:main} show three consistent trends across the memory, tool, and skill tracks. First, centralized access provides the strongest upper bound on task success. When the system is given an oracle resource owner or a global registry, performance is substantially higher than decentralized alternatives across all tracks. This confirms that the tasks are solvable when the required resources can be directly identified. However, centralized settings do not measure collaboration quality, since agents do not need to discover or negotiate access to remote resources through the network.

Second, fixed decentralized policies achieve moderate task success but incur nontrivial communication cost. Policies that always return content or results improve over metadata-only sharing because they expose more useful information to the requester. Always-forward policies further increase collaboration scores by expanding the search over the network, but this comes at a clear efficiency cost. The strongest fixed baseline is the policy that both returns content or results and forwards the request, suggesting that broad sharing and flooding can recover many required resources. However, its low efficiency indicates that this success is obtained through excessive communication rather than targeted collaboration. The budgeted fixed policy improves efficiency but reduces success, revealing the expected trade-off between resource discovery and communication cost.

Third, fully decentralized agents are more challenging to coordinate. The autonomous setting with only tool descriptions performs worst because agents must infer both the existence and usefulness of neighbor communication. Explicit network briefing improves task success and collaboration quality, showing that agents benefit from being told that useful resources may exist on neighboring nodes. Providing network metadata further improves success by helping agents route requests more effectively. Adding visited-node memory improves both success and efficiency, indicating that agents can avoid redundant exploration when they remember which nodes or resources have already been queried. Finally, imposing a communication budget increases efficiency but lowers success, suggesting that current autonomous agents still rely on relatively broad exploration to discover useful remote resources.

Across tracks, the skill setting is consistently harder than memory and tool use. This is expected because skills require not only locating the correct remote resource, but also understanding when and how to apply procedural knowledge. Overall, these results highlight the central challenge of decentralized agent collaboration: centralized registries offer high success but bypass collaboration, fixed policies provide useful but costly baselines, and fully decentralized agents must balance discovery, routing, privacy-preserving access, and communication efficiency.

\subsection{Discussion}

% \paragraph{Decentralized Network with Centralized Subnetwork (CDN)}

\paragraph{Robustness to Adversarial Agents}
Adversarial agents substantially degrade decentralized collaboration. When $k$ adversarial agents deliberately forward requests to the least promising neighbor, task success drops from 53.7 to 3.2 and collaboration quality decreases from 57.1 to 25.7. Efficiency also declines from 71.5 to 33.8. The degradation is most severe when adversarial agents occupy structurally or semantically important positions, such as hubs in scale-free networks or nodes connecting semantically related regions. These results show that decentralized collaboration is sensitive not only to the presence of adversarial behavior, but also to where adversarial agents are located in the network, see Table~\ref{tab:adv_agent}.

\begin{table}[t]
\centering
\small
\caption{Performance with adversarial agents. Adversarial agents forward requests to the least promising neighbor.}
\label{tab:adv_agent}
\begin{tabular}{lccc}
\toprule
\textbf{$k$} & \textbf{Success} & \textbf{Collaboration} & \textbf{Efficiency} \\
\midrule
0 (No Adv) & \textbf{53.7} & \textbf{58.1} & \textbf{71.5} \\
1    & 22.5 & 37.9 & 55.1 \\
2    & 13.5 & 30.1 & 41.6 \\
3    &  3.2 & 25.7 & 33.8 \\
% 0 (No Adv) & 76.8 & 78.3 & 83.8 \\
% 1    & 40.1 & 49.4 & 63.8 \\
% 2    & 17.5 & 25.7 & 35.8 \\
% 3    &  8.3 & 15.9 & 33.1 \\
\bottomrule
\end{tabular}
\end{table}

\paragraph{Robustness to Conflicting Information.}
We further study decentralized collaboration when agents receive conflicting routing information~\citep{cheng2026tmc,sun2026accurate}. We use a ring network so that each agent has exactly two neighbors, which provides a controlled setting for isolating conflict resolution. For each resource request, an agent receives two local knowledge sources that recommend where the request should be forwarded. In the no-conflict condition, both sources identify the neighbor that lies on a shortest path to the required resource owner. In the conflict condition, one source provides the correct direction while the other recommends the opposite neighbor. We randomize which source is correct across tasks so that agents cannot resolve conflicts based only on source identity. Introducing conflicting information reduces task success from 53.7 to 27.2, collaboration quality from 58.1 to 35.1, and efficiency from 71.5 to 30.8. The results suggest that decentralized collaboration degrades substantially when locally available routing evidence is inconsistent, even when the underlying network remains fully connected and the required resources remain reachable, see Table~\ref{tab:conflict}.

\begin{table}[t]
\centering
\small
\caption{Performance with and without conflicting information. Each agent receives two local routing sources. Under conflict, one source points toward the required resource owner while the other recommends the opposite direction.}
\label{tab:conflict}
\begin{tabular}{lccc}
\toprule
\textbf{Setting} & \textbf{Success} & \textbf{Collaboration} & \textbf{Efficiency} \\
\midrule
% No Conflicts   & 76.8 & 78.3 & 83.8 \\
% With Conflicts & 53.7 & 40.1 & 57.8 \\
No Conflicts & \textbf{53.7} & \textbf{58.1} & \textbf{71.5} \\
With Conflicts & 27.2 & 35.1 & 30.8 \\
\bottomrule
\end{tabular}
\end{table}

\section{Conclusion}

We introduce \benchmark, a benchmark for evaluating large-scale decentralized multi-agent collaboration. By formulating Graph-Constrained Dec-POMDP, DecArena evaluates whether agents can locate, access, and compose distributed resources without a centralized orchestrator. Our experiments show that even strong LLM agents struggle to proactively discover non-local resources using local information alone. These results suggest that decentralized collaboration remains a challenging task for existing agents to learn. We propose a strong baseline, \method, which is topology-aware, experience-adaptive and performs efficient and high quality collaboration. DecArena provides a controlled and scalable testbed for developing agents that can collaborate effectively in increasingly large and distributed agent ecosystems.

\section*{Limitations}
\benchmark intentionally isolates decentralized resource discovery and routing through synthetic, deterministic tool-calling tasks, which enables exact verification and controlled scaling but does not capture the full complexity of real-world collaboration involving ambiguous requests, heterogeneous APIs, natural-language expertise, changing permissions, execution failures, or strategic participants.

The headline setting also assumes connected networks, unique resource ownership, and reachable required resources, so failures measure discovery and coordination rather than network partition or resource unavailability.

Our evaluated topologies represent only several stylized network structures, and experiments at up to 1,000 agents do not fully reproduce the dynamics of web-scale agent ecosystems.

Finally, performance can depend on the underlying LLM, prompting strategy, network construction, and persistent routing state, while the current benchmark emphasizes resource discovery more than other decentralized challenges such as trust, authentication, incentive compatibility, asynchronous communication, and evolving ownership. Extending DecArena to real tools, dynamic networks, heterogeneous agents, and richer adversarial or privacy-sensitive settings is therefore an important direction for future work.

\section*{Ethical Considerations}
\benchmark is designed primarily as a controlled research benchmark and uses synthetic tools with deterministic execution rather than human data, reducing risks related to personal information, annotation labor, or sensitive real-world content. 
At the same time, advances in decentralized agent collaboration may have broader societal implications. Systems that can autonomously discover and invoke resources across organizational or ownership boundaries could improve scalability and enable capabilities to remain locally controlled, but they may also make distributed systems harder to audit and increase the risk of unauthorized resource access, information leakage, malicious routing, or propagation of incorrect information. The substantial degradation observed under adversarial agents and conflicting routing information further highlights these concerns.
Real-world deployment of such systems should therefore incorporate explicit authorization and consent, capability-level access control, provenance and logging, sandboxed execution, communication limits, and mechanisms for detecting malicious or unreliable participants. We view DecArena as a setting for studying these challenges rather than evidence that fully autonomous decentralized agent networks are ready for unrestricted deployment.

\bibliography{reference}

\appendix

\newpage

\section{\benchmark Benchmark Details}
\label{app:benchmark_details}

This section provides the full construction and evaluation details for \benchmark. The benchmark is designed to isolate a specific capability: whether agents can discover and use resources that are distributed across a communication network when no agent has access to a global resource registry or a global view of the network. We use synthetic tool calling as the primary task family because it provides exact, deterministic verification while preserving the central difficulty of decentralized resource discovery, multi-hop routing, and distributed execution.

\subsection{Tool Construction}
\label{app:tool-construction}

\paragraph{Tool bank.}
We construct a large synthetic bank containing more than 100k tools. The bank contains two primitive tool types, \emph{transform} and \emph{merge}. Tool identifiers are unique across the bank. Only a small fraction of the full tool bank is required by the benchmark tasks, while the remaining tools act as semantically plausible distractors.

The large tool space is intentional. If the environment contained only the tools required by the evaluation tasks, identifying a plausible tool description would already provide substantial information about the correct route. Instead, \benchmark places each required tool within a much larger capability ecosystem containing many irrelevant tools with similar metadata. The resulting problem therefore evaluates targeted resource discovery rather than enumeration over a small candidate set.

\paragraph{Tool types.}
A \emph{transform} consumes a single string and produces a single string,

\begin{equation}
y = f_t(x),
\end{equation} while a \emph{merge} consumes two or more strings and produces a single output,

\begin{equation}
y = f_t(x_1,\ldots,x_m), \qquad m \geq 2.
\end{equation}

Both tool types use the same underlying deterministic keyed transformation. They differ only in the number of inputs they consume.

Each tool is associated with a unique identifier, a semantic category, a natural-language description, an input schema, an output schema, and a private execution key. Tool identifiers are deliberately semantically structured. For example, an identifier may contain lexical cues associated with its category or function. These cues give an LLM a meaningful routing signal without revealing the tool output.

\paragraph{Private deterministic execution.}\label{app:deterministic}
Each tool $t$ receives a private secret key derived deterministically from a master seed and the tool identifier:

\begin{multline}
\mathrm{key}(t)=\mathrm{HMAC\text{-}SHA256}\\(\texttt{master\_seed},\texttt{tool\_id}_t).
\end{multline}

Given input strings $x_1,\ldots,x_m$, the execution procedure first canonicalizes the inputs by sorting them and concatenating them using a fixed delimiter. The canonicalized input is then processed using the private key. The resulting digest is Base32 encoded and truncated to a fixed-length output token. 

For a transform, $m=1$. For a merge, $m\geq2$. Therefore, the same deterministic construction supports both primitive types.

This design has four useful properties. First, repeated execution of the same tool on the same inputs always produces the same output. Second, all intermediate and final outputs can be reproduced exactly by the evaluator. Third, outputs cannot be reliably inferred from tool names, descriptions, or semantic categories because the execution key is private. Fourth, two semantically similar tools still produce unrelated outputs because their keys differ.

The private key remains with the tool owner. An agent that does not own a required tool cannot locally reconstruct its output. It must either obtain the result from the owner or, in resource-transfer variants of the benchmark, obtain the transferable resource itself.

\subsection{Task Construction}
\label{app:task-construction}

\paragraph{Multi-step pipelines.}
Each benchmark task defines a directed acyclic computation graph over synthetic tools. A task begins with a single raw input string and requires one or more tools to be executed in a prescribed dependency structure.

We organize tasks into three levels of compositional depth.

A level-1 task applies $k \in \{1,2,3\}$ transform tools independently to the raw input. These calls can be executed in parallel.

A level-2 task first executes the level-1 transforms and then applies one merge tool to their outputs.

A level-3 task extends the level-2 pipeline with one additional transform applied to the output of the merge.

For example:

\begin{verbatim}
Input: Z4P3

Step 1
  transform_green_river_00368(Z4P3)
  transform_amber_maple_00978(Z4P3)

Step 2
  merge_red_nova_00166(outputs from Step 1)

Step 3
  transform_blue_comet_00392
        (output from Step 2)

Return the output from Step 3.
\end{verbatim}

If $S_j$ denotes the set of tools at pipeline stage $j$, tools within the same stage may be executed independently, while stage $j+1$ becomes executable only after all required outputs from stage $j$ have been obtained. The benchmark therefore requires not only resource discovery but also preservation and composition of intermediate results.

\paragraph{Task families.}
Tasks are generated in families. A family fixes an input string together with a pool of tools used across multiple levels. Higher-level tasks extend lower-level tasks whenever the dependency structure permits.

For example, a level-2 instance may reuse the transforms from a corresponding level-1 instance and append a merge. A level-3 instance may further append a final transform. This construction allows compositional depth to vary while holding much of the underlying semantic content fixed.

As a result, differences between task levels primarily reflect the additional collaboration and dependency-management requirements rather than an unrelated change in tool semantics.

\paragraph{Evaluation task set.}
The evaluation set contains $140$ tasks: $60$ level-1 tasks, $40$ level-2 tasks, and $40$ level-3 tasks. Collectively, these tasks reference $186$ distinct required tools.

The number of benchmark tasks is intentionally much smaller than the size of the global tool bank. The benchmark therefore evaluates whether agents can locate a sparse set of relevant capabilities embedded in a much larger distributed resource space.

\paragraph{Required tools and distractors.}
For each benchmark environment, every tool required by the selected evaluation tasks is guaranteed to be present. The remaining resource slots are filled with distractor tools sampled from the global bank.

Distractors are fully valid tools. They have normal identifiers, descriptions, categories, schemas, and executable implementations, but they are not required by the current evaluation tasks. They are therefore indistinguishable from useful tools based only on their status in the environment.

The purpose of distractors is to prevent resource discovery from degenerating into simple category matching. Multiple tools may appear semantically related to the request even though only one exact tool produces the required output. As the tool bank grows beyond $100{,}000$ tools, the ratio of relevant to irrelevant resources becomes increasingly sparse, creating a large-scale discovery problem similar to realistic capability ecosystems.

\paragraph{Semantic categories.}
Tools are distributed across semantic categories. Category membership affects the semantic information available in tool metadata and, in the semantic network condition, influences both network structure and resource placement.

A configurable fraction of task families can be constructed so that their tools share a semantic category. The remaining families contain tools from multiple categories. This produces both semantically coherent tasks, for which routing toward a specialized region may be beneficial, and cross-domain tasks, for which the required resources may be distributed across different parts of the network.

Importantly, category information never uniquely identifies the required tool. It provides only a routing prior.

\subsection{Communication Networks and Resource Placement}
\label{app:network-construction}

For each benchmark environment, agents form an undirected communication graph $G=(V,E)$, where each node corresponds to an agent and an edge indicates that two agents may communicate directly. Every tool is assigned to exactly one agent. There is no resource replication unless explicitly introduced as a separate ablation.

We evaluate three network families: random, scale-free, and semantic. Each topology is paired with a corresponding resource-placement mechanism. This allows us to vary whether graph structure provides useful information about resource location.

\paragraph{Random networks.}
The random condition uses an undirected Erd\H{o}s--R\'enyi graph. For $N$ agents and target average degree $d$, each possible edge is sampled independently with probability

\begin{equation}
p=\min\left(\frac{d}{N-1},1\right).
\end{equation}

Unless otherwise specified, we use target average degree $d=6$.

Resource ownership is independent of network structure. Let $M$ denote the number of active tools placed in an environment. An initial resource count is independently sampled for each agent as

\begin{equation}
\tilde n_i \sim \mathrm{Poisson}\left(\frac{M}{N}\right).
\end{equation}

Because independent Poisson draws do not necessarily sum to exactly $M$, we apply an exact-total correction. If the sampled counts contain fewer than $M$ ownership slots, the missing slots are assigned to uniformly sampled agents. If they contain more than $M$, uniformly sampled ownership slots are removed. The final counts therefore satisfy

\begin{equation}
\sum_{i=1}^{N} n_i=M.
\end{equation}

The tools are then randomly permuted and partitioned according to these counts. Thus, neither graph position nor semantic category provides information about resource ownership in the random condition.

\paragraph{Scale-free networks.}
The scale-free condition uses preferential attachment. Let

\begin{equation}
m=\max\left(1,\min\left(\left\lfloor\frac{d}{2}\right\rfloor,N-1\right)\right).
\end{equation}

Construction begins from a clique of $m+1$ nodes. Each subsequently introduced agent creates $m$ edges to existing agents, with selection probability proportional to their current degree. With the default $d=6$, we use $m=3$.

This produces a heterogeneous degree distribution with a small number of highly connected hubs and many lower-degree agents.

Resource placement is also degree-biased. If $d_i$ is the realized degree of agent $i$, each tool is assigned to agent $i$ with probability

\begin{equation}
P(o(t)=i)=\frac{d_i^\alpha}{\sum_j d_j^\alpha},
\end{equation}

where we use $\alpha=1$ by default.

Consequently, structurally prominent agents are also more likely to own resources. This configuration models decentralized ecosystems in which well-connected agents tend to accumulate more capabilities or services.

\paragraph{Semantic networks.}
The semantic condition introduces correlation between agent specialization, network structure, and resource ownership.

% Each agent is first assigned one of the eight semantic categories. 
We initially connect all agents in a ring, which guarantees a connected backbone. Additional edges are then added until the target average degree is reached.

When constructing each additional edge, an initial endpoint is sampled uniformly. With probability $\rho$, the second endpoint is sampled from agents sharing the first endpoint's semantic category. Otherwise, it is sampled from the full agent population. We use

\begin{equation}
\rho=0.8.
\end{equation}

Resource placement uses a corresponding semantic bias. For a tool $t$ with category $c_t$, let

\begin{equation}
V(c_t)=\{i:c_i=c_t\}
\end{equation}

be the set of agents assigned to the same category. With probability $q$, the tool is assigned uniformly to an agent in $V(c_t)$. Otherwise, its owner is sampled uniformly from the entire network:

\begin{equation}
o(t)\sim
\begin{cases}
\mathrm{Uniform}(V(c_t)), & \text{with probability } q,\\
\mathrm{Uniform}(V), & \text{otherwise}.
\end{cases}
\end{equation}

We use $q=0.8$. If no agent represents category $c_t$, ownership falls back to uniform sampling.

The resulting graph contains semantic communities but does not provide an oracle route. A category may contain many agents, tools may be placed outside their matching communities, and a single task may require tools from multiple semantic regions.

\paragraph{Connectivity.}
All headline benchmark networks are connected. Random graph generation can occasionally produce multiple connected components, particularly at larger scales. We therefore compute the connected components after graph generation and, when necessary, connect each disconnected component to the largest component using one additional edge. If the initial graph contains $C$ components, this procedure introduces exactly $C-1$ edges, the minimum number required to produce a connected graph.

Scale-free and semantic graphs are connected by construction under the default generator, but the same connectivity check is applied to all topologies.

Connectivity guarantees that every required resource is in principle reachable from every entry agent. It does not reveal the path to that resource. Thus, failure reflects discovery or routing behavior rather than an impossible environment.

\paragraph{Unique resource ownership.}
Every active tool has exactly one owner. If $\mathcal{K}_i$ denotes the set of tools owned by agent $i$, then

\begin{equation}
\mathcal{K}_i \cap \mathcal{K}_j=\emptyset
\end{equation}

for all $i\neq j$, and

\begin{equation}
\bigcup_{i=1}^{N}\mathcal{K}_i=\mathcal{K}.
\end{equation}

This property is important because it prevents an agent from succeeding by accidentally encountering a duplicate copy of the required tool. A remote tool must be discovered through its actual owner.

\subsection{Local Observability}
\label{app:local-observability-details}

The defining constraint of \benchmark is local observability. No agent receives a global resource registry, the full adjacency matrix, or the mapping from tools to owners.

An agent has direct access only to its own resources and can communicate directly only with immediate graph neighbors. Information about a remote agent or resource must therefore propagate through a sequence of local interactions.

For agent $A_i$ at interaction step $t$, the available local information includes:
\begin{itemize}
    \item the request currently received by the agent
    \item currently available task inputs or intermediate outputs
    \item the tools locally owned by $A_i$
    \item the identities of the immediate neighbors of $A_i$
    \item messages and results previously received by $A_i$
    \item the within-task interaction history available under the evaluated agent design
    \item optional persistent local routing memory for methods that explicitly support cross-task adaptation
\end{itemize}

Only the entry agent initially receives the complete user task. Other agents learn about the task only through requests forwarded to them.

Most importantly, the environment does not expose a global function of the form

\begin{equation}
\texttt{lookup(tool)} \rightarrow \texttt{owner}.
\end{equation}

Locating a resource is therefore part of the benchmark rather than a preprocessing step performed by an external retriever.

\subsection{Request and Interaction Protocol}
\label{app:interaction-protocol}

A task begins at an entry agent. The entry agent may already own some of the required tools, while other required tools may be located several graph hops away.

When an agent receives a request, it chooses among the following primitive actions:

\begin{itemize}
    \item \texttt{Reject}: decline the request and return no result
    \item \texttt{Return}: execute an owned requested tool and return its output
    \item \texttt{Forward}: propagate the request to one or more immediate neighbors
    \item \texttt{Submit}: terminate the task and submit a final answer
\end{itemize}

\texttt{Return} and \texttt{Forward} may be selected simultaneously. For example, an agent may provide the result of one locally available resource while continuing to search for another required resource.

We deliberately do not prohibit inefficient actions. An agent may forward a request to several neighbors, revisit an earlier node, query an irrelevant resource, or continue searching after receiving a useful result. These behaviors are logged and penalized through the efficiency metrics rather than removed from the action space. This allows the benchmark to distinguish effective collaboration from success obtained through indiscriminate flooding.

\paragraph{Recursive decentralized routing.}
The default evaluation uses recursive routing. Once a request is forwarded to an intermediate agent, that agent independently chooses whether to answer, reject, or forward it to its own neighbors. The entry agent therefore does not centrally control every hop.

For example, if $A_1$ requests tool $t$ from neighbor $A_2$ and $A_2$ does not own $t$, $A_2$ may decide to forward the request to $A_5$. Agent $A_1$ need not know that $A_5$ exists before the request reaches it.

This distinction is central to the benchmark. The problem is not simply to choose a route over a known graph. The route itself emerges from a sequence of decisions made under different local observations.

\paragraph{Remote execution.}
The headline tool benchmark uses remote execution. The owner of a requested tool executes it locally and returns only the output. The underlying private key and tool implementation are not transferred.

If agent $A_j$ owns tool $t$, a successful remote call therefore has the form

\begin{equation}
A_i \rightarrow \cdots \rightarrow A_j:
\mathrm{request}(t,x),
\end{equation}

followed by

\begin{equation}
A_j \rightarrow \cdots \rightarrow A_i:
f_t(x).
\end{equation}

This models resources that depend on private data, credentials, proprietary code, or specialized execution environments and therefore cannot simply be copied to the requester.

\paragraph{Intermediate results.}
Multi-stage tasks require outputs from earlier tools to be reused as inputs to later tools. Returned results are therefore added to the task state and may be included in subsequent requests. A level-3 task may require several remote discoveries across multiple sequential stages before the final answer can be produced.

\subsection{Entry Agents and Distance-Based Difficulty}
\label{app:entry-distance}

For each task, we evaluate the same resource configuration from different entry agents. Let $\mathcal{T}(q)$ denote the tools required by task $q$, let $o(t)$ denote the owner of tool $t$, and let $d_G(\cdot,\cdot)$ denote shortest-path distance in the communication graph.

For entry agent $A_0$, we define task distance as

\begin{equation}
d(q,A_0)=\max_{t\in\mathcal{T}(q)} d_G(A_0,o(t)).
\end{equation}

The maximum captures the farthest required resource that must be reached to complete the full task.

We group runs into the following distance buckets:

\begin{itemize}
    \item \textbf{Local}: $d(q,A_0)=0$
    \item \textbf{1-hop}: $d(q,A_0)=1$
    \item \textbf{2-hop}: $d(q,A_0)=2$
    \item \textbf{3+-hop}: $d(q,A_0)\geq3$
\end{itemize}

Local cases measure whether an agent can correctly use resources it already owns. Increasing-distance cases progressively require more decentralized discovery and routing.

In the headline benchmark, the communication graph is connected and all task-required resources are present. Therefore, every task is theoretically solvable from every entry point. This allows success degradation with distance to be interpreted as a collaboration failure rather than lack of reachability.

\subsection{Exact Verification}
\label{app:exact-verification}

Because all synthetic tools are deterministic, the benchmark computes the complete gold execution trace for every task before evaluation. This includes the expected output of each required tool and the final task output.

Let $y^\star$ denote the gold final output and $\hat y$ the submitted output. End-to-end task success is

\begin{equation}
\mathrm{Success}=\mathbbm{1}[\hat y=y^\star].
\end{equation}

No learned evaluator is required.

The evaluator also records intermediate outputs. These diagnostics allow us to distinguish cases in which an agent found some required resources but failed to complete the full pipeline from cases in which it never reached the correct resources.

\subsection{Reproducibility}
\label{app:benchmark-reproducibility}

Tool generation, task generation, topology construction, and resource placement are seeded and deterministic given the benchmark configuration. Tool execution is deterministic by construction.

For network generation, topology sampling and resource placement use separate random-number streams. Given network seed $s$, graph construction uses seed $s$, while resource placement uses seed $s+1$. This prevents incidental changes in edge generation from changing the random sequence used for resource ownership.

All benchmark instances store the resulting graph adjacency, resource ownership, task specification, and generation configuration. The private execution information required for verification is retained by the environment but is not exposed to evaluated agents.

\subsection{Network Construction and Distributed Resource Ownership}
\label{app:network}

\method separates two aspects of the decentralized environment: the communication topology determines \emph{who can communicate with whom}, while resource placement determines \emph{where capabilities are located}. We construct these two components jointly so that different network families induce qualitatively different resource-discovery problems while preserving the same neighbor-to-neighbor interaction protocol.

\paragraph{Network configurations.}
For the tool benchmark, we construct networks with $N \in \{10, 50, 100, 500, 1000\}$ agents. For each network size, we instantiate three topology families: \textsc{Random}, \textsc{Scale-Free}, and \textsc{Semantic}. This results in 15 network configurations. Unless otherwise specified, all networks use target average degree $\bar d = 6$ and random seed $0$.

The full synthetic tool bank contains 100k+ tools, of which 186 distinct tools are referenced by the benchmark tasks. 
% We additionally sample an equal number of decoy tools, giving $M=372$ active tools in each network. The same active tool set is redistributed independently under each network family. 
Every tool has exactly one owner, so resource ownership forms a partition of the active tool set and no tool is replicated across agents.

The topology and ownership random variables are generated from separate deterministic random-number streams. The topology, agent categories, and connectivity repair use seed $s$, while resource placement uses seed $s+1$. This separation prevents incidental random choices made during graph construction from changing the resource allocation.

\subsubsection{Random Networks}

\paragraph{Communication topology.}
The random topology is an undirected Erd\H{o}s-R\'enyi graph. For $N>1$, every unordered pair of agents is connected independently with probability

\begin{equation}
p = \min\left(\frac{\bar d}{N-1}, 1\right).
\end{equation}

Thus, before connectivity repair, the expected degree of every agent is approximately $\bar d$. This topology deliberately contains no relationship between communication structure and resource semantics.

Because sparse Erd\H{o}s-R\'enyi graphs can contain disconnected components, we subsequently enforce connectivity. If the sampled graph contains $C>1$ connected components, we designate the largest component as the current main component and connect every remaining component to it with one additional undirected edge.
For each disconnected component, one endpoint is sampled from that component and the other from the accumulated main component.
The procedure adds exactly $C-1$ edges, which is the minimum number needed to make the graph connected.
The resulting change in average degree is therefore only

\begin{equation}
\Delta \bar d = \frac{2(C-1)}{N}.
\end{equation}

In practice, this repair is primarily relevant to the larger random graphs. The semantic and scale-free constructions are connected by construction.

\paragraph{Poisson resource allocation.}
Random networks also use a topology-independent resource distribution. Let $X_i$ denote the number of tools initially assigned to agent $i$. We first sample

\begin{equation}
\tilde X_i \sim \mathrm{Poisson}(\lambda),
\qquad
\lambda = \frac{M}{N},
\end{equation} independently for all agents. This produces natural heterogeneity in local resource capacity rather than forcing every agent to own exactly the same number of tools.

Because independent Poisson draws do not necessarily sum exactly to $M$, we apply an exact-total correction.
Let

\begin{equation}
\Delta = M - \sum_{i=1}^{N} \tilde X_i.
\end{equation}

If $\Delta>0$, the missing $\Delta$ ownership slots are assigned to uniformly sampled agents.
If $\Delta<0$, $|\Delta|$ existing ownership slots are sampled uniformly and removed.
This yields final counts $X_i$ satisfying

\begin{equation}
\sum_{i=1}^{N} X_i = M.
\end{equation}

Finally, the $M$ tool identifiers are randomly permuted and partitioned according to these counts.
Consequently, random placement has no dependence on either tool category or graph degree.
The distribution should therefore be interpreted as a Poisson-like heterogeneous allocation with an exact-total correction, rather than an unconstrained collection of independent Poisson random variables.

For the tool benchmark, the expected number of tools per agent is therefore $37.2$, $7.44$, $3.72$, $0.744$, and $0.372$ for $N=10$, $50$, $100$, $500$, and $1000$, respectively.
At the largest scales, many agents intentionally own no local tool.
Such nodes remain useful as communication intermediaries and make resource discovery distinct from simply querying every node for its local capability.

\subsubsection{Scale-Free Networks}

\paragraph{Communication topology.}
The scale-free topology follows preferential attachment.
We use

\begin{equation}
m = \max\left(1, \min\left(\left\lfloor\frac{\bar d}{2}\right\rfloor, N-1\right)\right)
\end{equation}

attachment edges per newly introduced node.
With the default $\bar d=6$, this gives $m=3$ for all benchmark network sizes.

Construction begins with a clique of $m+1$ agents.
Each subsequent agent connects to $m$ distinct existing agents.
The probability of choosing existing agent $j$ is proportional to its current degree,

\begin{equation}
P(j \mid G_t)
=
\frac{d_j(t)}
{\sum_{\ell} d_\ell(t)},
\end{equation}

up to a negligible numerical smoothing constant used to avoid zero probability during implementation.
Preferential attachment creates a heterogeneous degree distribution containing a small number of high-degree hubs and many lower-degree peripheral agents.
The expected average degree approaches $2m \approx 6$ as $N$ grows.

\paragraph{Degree-biased resource ownership.}
Scale-free networks intentionally couple structural centrality with resource availability.
After the graph is generated, tool $k$ is assigned independently to agent $i$ with probability

\begin{equation}
P(o(k)=i)
=
\frac{d_i^{\alpha}}
{\sum_{j=1}^{N} d_j^{\alpha}},
\end{equation}

where $d_i$ is the realized degree of agent $i$ and $\alpha$ controls the strength of the ownership bias.
We use $\alpha=1$ in the benchmark.

Thus, high-degree agents are not only easier to reach but also more likely to own resources.
This models ecosystems in which highly connected or prominent agents accumulate more capabilities, services, or expertise.
Unlike the random construction, the number of resources owned by an agent is therefore positively correlated with its network centrality.

The construction creates two competing effects for decentralized routing.
Hubs provide short paths and concentrate useful resources, which can make successful routes efficient.
At the same time, identifying the appropriate hub from only local information is non-trivial, and poor routing decisions around high-degree nodes can induce substantial fan-out.

\subsubsection{Semantic Networks}

\paragraph{Agent semantic categories.}
Semantic networks introduce local alignment between graph structure and capability semantics.
Each agent is first assigned one of the eight tool categories.
Agent categories are sampled independently and uniformly from the categories represented in the active tool collection.
Let $c_i$ denote the category assigned to agent $i$.

\paragraph{Communication topology.}
We first connect all agents in a ring.
This guarantees connectivity while assigning every agent at least two structural neighbors before additional semantic edges are added.

We then add undirected edges until the graph reaches

\begin{equation}
\small
|E|
=
\min\left(
\left\lfloor
\frac{N \min(\bar d,N-1)}{2}
\right\rfloor,
\frac{N(N-1)}{2}
\right).
\end{equation}

For every candidate edge, an endpoint $i$ is sampled uniformly.
With probability $\rho$, the second endpoint is sampled from another agent sharing the same category as $i$.
Otherwise it is sampled from the complete agent set.
Self-loops and duplicate edges are rejected.
We use

\begin{equation}
\rho = 0.8.
\end{equation}

Consequently, most new edges are biased toward semantically similar agents, while a smaller fraction of cross-category edges maintains bridges between semantic regions of the network.
Because the graph begins with a ring and then adds edges until the target edge count is reached, its realized average degree is exactly $6$ for the benchmark sizes considered here, except for trivial finite-size restrictions.

\paragraph{Semantic resource allocation.}
Tool placement uses the same semantic categories.
For a tool $k$ with category $c_k$, let

\begin{equation}
A(c_k) = \{i : c_i = c_k\}
\end{equation}

denote agents assigned to that category.
With probability $q$, the tool is assigned uniformly to an agent in $A(c_k)$.
With probability $1-q$, it is assigned uniformly to any agent in the network:

\begin{equation}
\small
o(k)
\sim
\begin{cases}
\mathrm{Uniform}(A(c_k)), & \text{with probability } q, \\
\mathrm{Uniform}(\{1,\ldots,N\}), & \text{with probability } 1-q.
\end{cases}
\end{equation}

We set $q=0.8$.
If no agent happens to receive the relevant semantic category, placement falls back to uniform assignment.

This construction creates a noisy but useful correspondence between network locality and capability semantics.
An agent seeking a tool from a particular category can benefit from routing toward semantically related neighborhoods, but category information is not an oracle for exact resource ownership.
First, 20\% of tools are placed without semantic matching.
Second, multiple agents can share a category.
Third, tasks frequently require several tools, potentially from different categories.
Semantic routing therefore provides a useful inductive bias without collapsing the problem into deterministic category lookup.

\subsubsection{Comparison of the Three Network Families}

The three constructions independently stress different aspects of decentralized collaboration.
Random networks provide a neutral condition in which neither graph structure nor ownership conveys semantic routing information.
Scale-free networks introduce strong structural heterogeneity and correlate resource ownership with degree.
Semantic networks instead introduce local homophily between capability categories and both communication edges and resource ownership.

Importantly, these conditions modify only the environment.
The agent interaction protocol is identical across all three settings.
An agent observes its own local resources and immediate neighbors, but receives neither a global adjacency matrix nor a global resource registry.
A distant resource can therefore be discovered only through a sequence of neighbor-to-neighbor interactions.

\subsubsection{Realized Network Statistics}
\label{app:network_stats}

Table~\ref{tab:network_stats} reports the realized statistics obtained with the benchmark seed and the current network builder.
For each configuration, we report the number of edges, degree distribution, and resource-count distribution.
The degree statistics are computed after connectivity repair.
Resource counts always sum to $M=372$.

Several expected patterns are visible.
First, random and semantic networks maintain mean degree close to the target value of $6$ across all scales.
Second, scale-free networks have similar mean degree but substantially heavier upper tails: the maximum degree increases from $8$ at $N=10$ to $95$ at $N=1000$.
Third, degree-biased ownership produces correspondingly concentrated resource allocation in scale-free networks.
For example, with $N=100$, the most resource-rich scale-free agent owns 19 tools, compared with 9 in the random network and 10 in the semantic network.
Finally, as $N$ becomes larger than the number of active tools, zero-resource agents become increasingly common.
This is intentional: large decentralized systems may contain many routing or intermediary agents that do not themselves own a capability relevant to the current benchmark task.

\begin{table*}[t]
\centering
\caption{Realized network statistics for the tool benchmark using seed $0$, target average degree $\bar d=6$, $M=372$ active tools, scale-free ownership exponent $\alpha=1$, semantic edge probability $\rho=0.8$, and semantic ownership probability $q=0.8$. Degree and item-count columns report mean followed by $[\min,\mathrm{P50},\mathrm{P95},\max]$.}
\label{tab:network_stats}
% \resizebox{\linewidth}{!}{
\begin{tabular}{llrcc}
\toprule
\textbf{Agents} & \textbf{Topology} & \textbf{Edges} &
\textbf{Degree} &
\textbf{Tools per Agent} \\
\midrule
10
& Random
& 31
& 6.20 $[4,6,8,8]$
& 37.20 $[31,36,44,44]$ \\
& Scale-Free
& 24
& 4.80 $[3,4,7,8]$
& 37.20 $[17,36,55,62]$ \\
& Semantic
& 30
& 6.00 $[3,6,8,8]$
& 37.20 $[27,32,55,59]$ \\
\midrule
50
& Random
& 144
& 5.76 $[3,5,10,11]$
& 7.44 $[3,7,11,13]$ \\
& Scale-Free
& 144
& 5.76 $[3,4,15,19]$
& 7.44 $[0,6,18,25]$ \\
& Semantic
& 150
& 6.00 $[3,6,9,9]$
& 7.44 $[1,6,17,21]$ \\
\midrule
100
& Random
& 300
& 6.00 $[1,6,10,12]$
& 3.72 $[1,3,7,9]$ \\
& Scale-Free
& 294
& 5.88 $[3,4,14,24]$
& 3.72 $[0,3,11,19]$ \\
& Semantic
& 300
& 6.00 $[3,6,9,13]$
& 3.72 $[0,3,8,10]$ \\
\midrule
500
& Random
& 1482
& 5.93 $[1,6,10,16]$
& 0.744 $[0,1,2,4]$ \\
& Scale-Free
& 1494
& 5.98 $[3,4,16,59]$
& 0.744 $[0,0,3,10]$ \\
& Semantic
& 1500
& 6.00 $[2,6,9,12]$
& 0.744 $[0,1,2,5]$ \\
\midrule
1000
& Random
& 3016
& 6.03 $[1,6,10,14]$
& 0.372 $[0,0,2,4]$ \\
& Scale-Free
& 2994
& 5.99 $[3,4,15,95]$
& 0.372 $[0,0,2,9]$ \\
& Semantic
& 3000
& 6.00 $[2,6,9,14]$
& 0.372 $[0,0,2,3]$ \\
\bottomrule
\end{tabular}
% }
\end{table*}

\paragraph{Path-length characteristics.}
The graph families also differ in the routes they induce.
Table~\ref{tab:network_path_stats} reports diameter and average shortest-path length for the same realized graphs.
At larger scales, scale-free networks yield shorter paths because hubs connect otherwise distant portions of the graph.
For example, at $N=1000$, the scale-free graph has diameter $6$ and mean shortest-path length $3.43$, compared with diameter $9$ and mean shortest-path length $4.06$ for the random graph and diameter $8$ and mean shortest-path length $4.14$ for the semantic graph.
Thus, scale-free networks can make resources structurally closer while simultaneously creating a more heterogeneous routing landscape.

\begin{table}[t]
\centering
\caption{Shortest-path statistics of the realized communication graphs.}
\label{tab:network_path_stats}
\resizebox{\linewidth}{!}{
\begin{tabular}{llcc}
\toprule
\textbf{Agents} & \textbf{Topology} &
\textbf{Diameter} &
\textbf{Avg. Shortest Path} \\
\midrule
10   & Random     & 2 & 1.31 \\
     & Scale-Free & 2 & 1.47 \\
     & Semantic   & 2 & 1.33 \\
50   & Random     & 4 & 2.38 \\
     & Scale-Free & 4 & 2.29 \\
     & Semantic   & 4 & 2.39 \\
100  & Random     & 5 & 2.76 \\
     & Scale-Free & 5 & 2.66 \\
     & Semantic   & 5 & 2.79 \\
500  & Random     & 7 & 3.70 \\
     & Scale-Free & 5 & 3.23 \\
     & Semantic   & 6 & 3.75 \\
1000 & Random     & 9 & 4.06 \\
     & Scale-Free & 6 & 3.43 \\
     & Semantic   & 8 & 4.14 \\
\bottomrule
\end{tabular}
}
\end{table}

\paragraph{Reachability versus routing difficulty.}
Connectivity is enforced so that benchmark failure cannot be attributed simply to the absence of a communication path.
Every active tool has exactly one owner, and every owner is reachable from every possible entry agent.
However, connectivity does not make discovery trivial.
Agents do not observe global shortest paths, remote resource locations, or the full graph.
As the number of agents increases, a required tool may be several hops away, and many intermediate agents may own no relevant resource.
The benchmark therefore evaluates whether an agent can discover and exploit the existing path using only local routing decisions rather than whether such a path exists.

\subsection{Evaluation Metrics}
\label{app:metrics}

We distinguish end-to-end task success from partial progress. This avoids treating a pipeline with one correct intermediate tool output as a successful task.

\paragraph{Notation.}
For one task, let $T$ denote the set of required tools. Let $d_t$ be the shortest-path distance from the entry agent to the owner of tool $t$. The set of required remote tools is
\begin{equation}
R=\{t \in T : d_t>0\}.
\end{equation}
Let $\mathcal{V}=(v_1,\ldots,v_N)$ be the node-visit trajectory and let $N=|\mathcal{V}|$. Let $K$ be the set of scored intermediate-output keys. Each $k \in K$ has gold value $g_k$ and submitted value $\hat{s}_k$. The mapping $\tau:K\rightarrow T$ associates each output key with the tool that produced it. Let $Q$ be the set of tools queried during the run and let $\Gamma$ be the set of tools for which a result was returned.

\paragraph{End-to-end task success.}
Let $y^\star$ be the gold final output and $\hat{y}$ the submitted final output. Case-level success is binary:
\begin{equation}
\mathrm{Success}=\mathbbm{1}[\hat{y}=y^\star].
\end{equation}
Dataset-level success is the mean of the case-level indicator and is reported as a percentage.

\paragraph{Per-tool output accuracy.}
For each scored tool output, define
\begin{equation}
a_t=\mathbbm{1}[\hat{s}_k=g_k], \qquad t=\tau(k).
\end{equation}
The mean per-tool accuracy is
\begin{equation}
\mathrm{ToolAcc}=\frac{1}{|\tau(K)|}\sum_{t\in\tau(K)}a_t.
\end{equation}
If no intermediate outputs are available for scoring, this diagnostic is undefined. Tool accuracy is reported as partial-credit information and is not used as a substitute for end-to-end success.

\paragraph{Collaboration quality.}
For tasks with at least one remote required tool, we measure whether the run queried the right remote capabilities, obtained their results, and used them correctly:
\begin{equation}
\mathrm{QueryRate}=\frac{|R\cap Q|}{|R|}.
\end{equation}
\begin{equation}
\mathrm{ResponseRate}=\frac{|R\cap \Gamma|}{|R|}.
\end{equation}
\begin{equation}
\mathrm{CorrectUseRate}=\frac{|\{t\in R:a_t=1\}|}{|R|}.
\end{equation}
The aggregate collaboration score is
\begin{multline}
    \mathrm{Collaboration}=(\mathrm{QueryRate}+\mathrm{ResponseRate}\\
                           +\mathrm{CorrectUseRate})/3.
\end{multline}
For tasks with $R=\varnothing$, collaboration is not required. The evaluator assigns these cases a collaboration score of $100$ by convention. We therefore also report collaboration over remote tasks only, which prevents local-only cases from inflating the interpretation of decentralized collaborative ability.

\paragraph{Visit efficiency.}
The current evaluator uses the distance-based reference
\begin{equation}
V_{\mathrm{lb}}=1+\max_{t\in T}d_t.
\end{equation}
This quantity is a lower bound on the number of sequential node positions needed to reach the farthest required owner. It is not the exact optimal number of node visits when required tools lie on different graph branches. We therefore use it only as a normalization reference:
\begin{equation}
\mathrm{VisitEff}=100\cdot\frac{V_{\mathrm{lb}}}{\max(V_{\mathrm{lb}},N,1)}.
\end{equation}
This metric rewards compact trajectories while remaining simple to compute. When comparing systems with substantially different branching patterns, raw visited-node counts should be reported alongside it.

\paragraph{Query focus.}
Let $F$ be the set of visits that forward to at least one neighbor. For each $v\in F$, let $f_v$ be the number of neighbors selected at that visit. Define the mean forwarding fan-out as
\begin{equation}
\bar{f}=\frac{1}{|F|}\sum_{v\in F}f_v.
\end{equation}
When $F=\varnothing$, we set $\bar{f}=1$. Query focus is
\begin{equation}
\mathrm{QueryFocus}=\frac{100}{\bar{f}}.
\end{equation}
A score of $100$ corresponds to single-neighbor forwarding whenever forwarding occurs. Lower scores indicate broader fan-out. Because narrow routing can be efficient but incorrect, this metric is interpreted jointly with success and collaboration quality.

\paragraph{Round efficiency.}
Partition the required tools into pipeline stages $\{S_1,\ldots,S_m\}$. Under parallel breadth-first propagation within each stage, we use the following idealized reference round count:
\begin{equation}
R_{\mathrm{bfs}}=1+\sum_{j=1}^{m}2\max_{t\in S_j}d_t.
\end{equation}
The factor of two accounts for request propagation to the farthest required owner in a stage and return of the result. Let
\begin{equation}
R_{\mathrm{used}}=\max_{v\in\mathcal{V}}\mathrm{round}(v).
\end{equation}
Then
\begin{equation}
\small
\mathrm{RoundEff}=100\cdot\min\left(1,\frac{R_{\mathrm{bfs}}}{\max(1,R_{\mathrm{used}})}\right).
\end{equation}
This is a reference-normalized score rather than an exact shortest schedule for arbitrary asynchronous execution.

\paragraph{Aggregate efficiency.}
The aggregate efficiency score is the mean of the defined submetrics:
\begin{multline}
\small
\mathrm{Efficiency}=\operatorname{mean}(\mathrm{VisitEff},\mathrm{QueryFocus},\\\small\mathrm{RoundEff}).
\end{multline}
We report both unconditional efficiency and efficiency conditioned on successful completion. The conditional score is important because a failed run may stop early and appear artificially cheap.

\subsection{Additional Diagnostics}
\label{sec:diagnostics}

The benchmark records additional unscored statistics to support failure analysis. These include total token usage, wall-clock time, the number of repeated node visits, and rejection behavior.

Let \texttt{num\_repeated\_nodes} denote the number of repeated node visits. We report
\begin{equation}
\mathrm{RepeatRate}=\frac{\texttt{num\_repeated\_nodes}}{N}.
\end{equation}
A visit is \emph{ask-owned} if the request targets a tool owned by the receiving agent and \emph{ask-unowned} otherwise. We then define
\begin{equation}
\small
\mathrm{RefusalRate}=\frac{\#\{\mathrm{ask\text{-}owned}\land\texttt{Reject}\}}{\#\{\mathrm{ask\text{-}owned}\}}.
\end{equation}
\begin{equation}
\small
\mathrm{DeadEndRate}=\frac{\#\{\mathrm{ask\text{-}unowned}\land\texttt{Reject}\}}{\#\{\mathrm{ask\text{-}unowned}\}}.
\end{equation}
A rate is left undefined when its denominator is zero.

Together, these diagnostics distinguish several sources of failure. An agent may fail to identify the required capability, route toward the correct owner, obtain a response, preserve intermediate results, or synthesize the final pipeline output. Reporting these diagnostics alongside success, collaboration, and efficiency helps separate routing quality from favorable resource placement or premature termination.

\section{Details of Evaluated Agents}
\label{app:agents}

We compare several agent designs that differ in the information available to the policy and in how routing decisions are produced or adapted over time. Unless otherwise specified, all agents use the same base LLM, local observation space, structured action space, and neighbor-to-neighbor communication protocol. In particular, changing the agent design does not change tool ownership, network connectivity, or the underlying task.

At a visit to agent $A_i$ at round $t$, the agent observes its current local state $o_{i,t}$, including its locally owned tools, immediate neighbors, received requests or replies, and task information visible at that node. The policy then selects among the benchmark actions \texttt{Reject}, \texttt{Return}, \texttt{Forward}, and, for the entry agent, \texttt{Submit\_Final}. Except for the explicitly privileged oracle baselines, agents do not observe the global graph or the mapping from tools to owners.

\subsection{Reflex Agent}
\label{sec:agent-reflex}

The Reflex agent is the simplest LLM-based decentralized baseline. It makes each decision solely from the information currently visible at the node:

\begin{equation}
u_{i,t}\sim\pi_{\mathrm{LLM}}\left(\cdot\mid o_{i,t}\right).
\end{equation}

The agent does not maintain an explicit memory of earlier decisions. In particular, its prompt does not include a stored history of previously contacted nodes, previous forwarding actions, or outcomes from earlier requests. Any information available to the policy must therefore be present in the current observation.

This baseline tests whether a pretrained LLM can infer useful decentralized routing behavior directly from local context. For example, if an agent does not own a requested tool, it must decide whether to reject the request or forward it to one of its immediate neighbors without access to a learned routing table or cross-task experience.

\subsection{Reflex + History}
\label{sec:agent-history}

Reflex + History augments the Reflex agent with an explicit record of its previous interactions. Its decision can therefore condition on a local interaction history $h_{i,t}$:

\begin{equation}
u_{i,t}\sim\pi_{\mathrm{LLM}}\left(\cdot\mid h_{i,t}\right).
\end{equation}

For each previous visit, the history records the interaction round, agents from which messages were received, requested tool identifiers, selected actions, forwarding destinations, tools returned or executed, and execution or parsing errors when applicable. The history is rendered into the LLM prompt before the next decision.

The purpose of this baseline is to isolate the value of explicit interaction memory. In particular, an agent can use its history to avoid repeatedly forwarding the same request to the same neighbor, recognize that a previous route did not return a useful result, or retain information about earlier successful interactions.

Unlike the experience-conditioned methods below, Reflex + History does not perform explicit retrieval or learn a routing value function. It simply exposes the recorded interaction trace to the LLM and allows the model to reason over that trace directly.

\subsection{Collaborative Prompting}
\label{sec:agent-collaborative-prompting}

Collaborative Prompting tests whether poor decentralized performance is primarily caused by LLMs being insufficiently willing to seek help. It uses the same observation, memory, and action space as the Reflex agent, but strengthens the system prompt with explicit collaboration instructions.

The prompt encourages an agent to help whenever possible, to return a local result when it owns a useful tool, and to forward a request when a neighbor may plausibly help. In particular, the policy is instructed to strongly prefer collaboration over rejection and to avoid \texttt{Reject} unless neither the current agent nor any plausible neighbor can contribute.

Importantly, Collaborative Prompting does not specify which neighbor should be selected. The LLM must still infer the forwarding destination from its local observation:

\begin{equation}
u_{i,t}\sim\pi_{\mathrm{LLM}}\left(\cdot\mid o_{i,t},p_{\mathrm{collab}}\right),
\end{equation}

where $p_{\mathrm{collab}}$ denotes the additional collaboration instruction.

Thus, this baseline changes the agent's behavioral incentive but provides no additional information about resource locations or network topology. It distinguishes failures caused by unwillingness to collaborate from failures caused by an inability to identify a useful route.

\subsection{Broadcast}
\label{sec:agent-broadcast}

Broadcast provides a high-recall decentralized search baseline. The LLM is first queried using the standard local policy. The resulting action is then post-processed so that, whenever the agent has neighbors, the request is forwarded to all immediate neighbors.

More specifically, if forwarding is possible, \texttt{Reject} is removed from the final action set and \texttt{Forward} is added. For every tool-specific request being propagated, an equivalent request is created for each immediate neighbor. If the LLM has already produced a valid forwarding request, its requested tool, input, and request type are reused. Otherwise, the forwarding request is reconstructed from the current incoming request.

Broadcast does not prevent useful local computation. An agent may simultaneously execute an owned tool and return its result while forwarding the request to its neighbors. The distinction from the other agents is that neighbor selection is no longer selective:

\begin{equation}
F_{i,t}=\mathcal{N}_i,
\end{equation}

whenever forwarding is applicable, where $\mathcal{N}_i$ is the immediate-neighbor set of agent $A_i$.

Broadcast therefore approximates exhaustive graph search while retaining the same local communication constraint. It provides a useful reference for the maximum resource coverage obtainable through aggressive decentralized exploration, while typically incurring substantially greater communication cost, forwarding fan-out, and redundant node visits.

\subsection{Oracle Route}
\label{sec:agent-oracle-route}

Oracle Route provides privileged routing information to the entry agent while leaving all other agents unchanged. For every tool required by the task, the benchmark computes a shortest path from the entry agent to an owner of that tool.

The entry agent receives an additional routing section containing, for each required tool:

\begin{itemize}
    \item the tool identifier
    \item its owner
    \item a shortest path from the entry agent to that owner
    \item the first neighbor on that path
\end{itemize}

If a required tool is already owned by the entry agent, the hint explicitly identifies it as local.

Let $o(t)$ denote the owner of required tool $t$ and let

\begin{equation}
P_t=(A_0,A_{p_1},\ldots,o(t))
\end{equation}

be a shortest path from entry agent $A_0$ to that owner. Oracle Route reveals $P_t$ to $A_0$, including the first-hop recommendation $A_{p_1}$.

Only the entry agent receives this privileged information. Intermediate agents continue to make their own decisions from normal local observations. Oracle Route therefore removes uncertainty about initial resource location and route selection while preserving decentralized behavior after the first forwarding step. It serves as a diagnostic upper bound on how much performance can improve when the entry agent knows where the required resources are located.

\subsection{Oracle Route All}
\label{sec:agent-oracle-route-all}

Oracle Route All is a stronger privileged baseline. It constructs the same oracle routing information as Oracle Route, but exposes this information to every agent that participates in the task.

For each required tool, all visited agents can observe the owner and the shortest path computed from the original entry agent to that owner. Thus, an intermediate agent that lies on the supplied route can directly identify the next node on the oracle path rather than rediscovering the route from its ordinary local context.

The oracle information is not recomputed into a new globally optimal route at every intermediate node. Instead, all agents receive the same entry-originated shortest-path description. This matches the implementation and preserves a single oracle reference route for each required resource.

Oracle Route All therefore removes most of the resource-location and multi-hop routing uncertainty while retaining the benchmark's edge-constrained communication mechanism. Messages must still traverse actual graph edges, tools remain locally owned, and remote tools must still be executed by their owners. The baseline is intended as a routing oracle rather than a realistic decentralized agent.

\subsection{Experience-Conditioned Agent}
\label{sec:agent-experience}

The Experience-Conditioned agent augments the LLM with persistent, agent-local experience accumulated from previous interactions. Unlike Reflex + History, which exposes a raw interaction trace, this method stores structured evidence about successful resource retrieval.

Each agent maintains its own experience memory. When a reply for tool $t$ is observed from agent $A_j$, the receiving agent records $A_j$ as a successful helper for $t$. Repeated observations accumulate a success count. For returned tool results, the agent additionally caches the exact association between tool identifier, tool input, returned output, and observed helper. Returned transferable content can be cached analogously.

For a future request, only experience relevant to the currently requested tools is retrieved. The memory presented to the LLM contains, when available:

\begin{itemize}
    \item previously successful helpers for the requested tool
    \item the number of successful observations associated with each helper
    \item whether a helper is currently an immediate neighbor
    \item exact cached results for matching tool-input pairs
    \item previously observed transferable tool content
    \item request-level state indicating whether the current request was previously seen, forwarded, or rejected
\end{itemize}

The implementation retains at most the five highest-count helper entries and up to five matching cached results per tool when rendering relevant experience.

The resulting policy can be written as

\begin{equation}
u_{i,t}\sim\pi_{\mathrm{LLM}}\left(\cdot\mid h_{i,t},\mathcal{H}_i^{+}\right),
\end{equation}

where $\mathcal{H}_i^{+}$ contains only experience locally observed by agent $A_i$. No global trajectory database or centralized resource map is constructed.

This baseline evaluates whether decentralized routing improves simply by remembering which agents previously helped with similar tools. It is particularly useful for distinguishing semantic generalization from explicit location memorization. A successful helper is evidence about a useful direction in the network, but it need not be the final tool owner and may no longer be directly reachable from the current local state.

\subsection{Numerical Q-Routing}
\label{sec:q-routing-details}

We include an online numerical routing baseline inspired by Q-Routing~\citep{boyan1993packet}. Each agent maintains its own routing table indexed by tool and immediate neighbor. The table estimates the value of forwarding a request for tool $t$ from agent $A_i$ to neighbor $A_j$.

For each tool-neighbor pair, the implementation maintains two quantities. The first, $\mathrm{resp}_i(t,j)$, measures whether requests for tool $t$ have successfully produced replies through neighbor $A_j$. The second estimates the distance from $A_i$ to the tool through $A_j$. The corresponding hop-efficiency term is

\begin{equation}
\mathrm{eff}_i(t,j)=\frac{1}{D_i(t,j)}.
\end{equation}

The routing value exposed to the LLM is

\begin{equation}
Q_i(t,j)
=
\frac{1}{2}
\left(
\mathrm{resp}_i(t,j)
+
\mathrm{eff}_i(t,j)
\right).
\end{equation}

All unseen tool-neighbor pairs are initialized to zero.

\paragraph{Online update.}
Learning occurs when an agent observes a reply for tool $t$ arriving from an immediate neighbor $A_j$. Receiving a reply provides direct evidence that forwarding through $A_j$ can reach a responder. The response estimate is therefore updated toward one:

\begin{equation}
\mathrm{resp}_i(t,j)
\leftarrow
(1-\alpha)\mathrm{resp}_i(t,j)+\alpha,
\end{equation}

where the default learning rate is $\alpha=0.2$.

The distance estimate uses a local bootstrap. Let $D_j^\star(t)$ denote neighbor $A_j$'s best currently known distance to tool $t$. If $A_j$ owns the tool, then

\begin{equation}
D_j^\star(t)=0.
\end{equation}

Otherwise, $D_j^\star(t)$ is the smallest previously successful distance estimate available at $A_j$, excluding a route immediately back through $A_i$. Agent $A_i$ uses the target

\begin{equation}
\hat D_i(t,j)=1+D_j^\star(t).
\end{equation}

For an existing estimate,

\begin{equation}
D_i(t,j)
\leftarrow
(1-\alpha)D_i(t,j)
+
\alpha\hat D_i(t,j).
\end{equation}

On the first grounded observation, the target distance is assigned directly.

If the neighbor does not own the requested tool and has no grounded downstream distance estimate, the response component is still updated because a reply was observed, but the distance component is left unchanged. If no reply is ever observed through a neighbor, the corresponding routing value remains at its initial value.

\paragraph{Decentralized routing state.}
Every agent owns a separate Q-table. Before each decision, the entries corresponding to the currently relevant tool are rendered into that agent's prompt as a ranking over its immediate neighbors. The LLM then selects an action using the numerical routing estimates together with its ordinary local observation.

The implementation uses a shared runtime registry to access the replying neighbor's distance estimate during the bootstrap update. This registry is used only for the estimate of the immediate neighbor from which the reply was observed. It does not expose the full network topology, remote resource locations, or arbitrary non-neighbor routing tables to the policy. Conceptually, it is equivalent to returning the neighbor's scalar routing estimate together with the local reply.

The routing table can be updated continuously or frozen after a specified number of training tasks. Under the frozen condition, subsequent evaluation tasks can read the learned routing values but cannot modify them. This permits a clean separation between online adaptation and evaluation of a fixed learned routing state.

\subsection{Comparison of Agent Information}
\label{sec:agent-information-comparison}

The evaluated agents can be understood by the additional information introduced beyond the current local observation. Reflex receives no additional state. Reflex + History exposes previous interactions. Collaborative Prompting changes only the behavioral instruction. Broadcast removes selective neighbor choice by forcing full local fan-out. Experience-Conditioned routing retrieves structured evidence from previous successful interactions. Numerical Q-Routing maintains an explicit learned neighbor-value table. Oracle Route and Oracle Route All introduce privileged ground-truth routing information and are included only as diagnostic upper bounds.

Crucially, except for the oracle baselines, none of these methods receives the global mapping from tools to owners or a globally searchable representation of the communication graph. Differences in performance therefore reflect different mechanisms for using local observations, communication history, and accumulated routing experience rather than access to a centralized resource registry.

\subsection{Training Learned Collaboration Policies}
\label{sec:training-details}

We study whether decentralized collaboration can be improved through policy post-training. We consider three complementary approaches: supervised fine-tuning (SFT), on-policy self-distillation (OPSD), and group relative policy optimization (GRPO). SFT learns from previously collected successful collaboration trajectories. OPSD uses a stronger teacher to supervise states generated by the current student policy. GRPO directly optimizes the policy using rewards obtained from online interaction with the decentralized environment.

All three methods train a single shared policy that is used by every agent in the network. The model is therefore not specialized to a particular node identity. At each node visit, the shared policy receives only the local observation available to the acting agent and predicts the structured decision consumed by the environment. This parameter sharing allows a policy trained on small networks to be deployed to previously unseen agents and substantially larger communication graphs.

\subsubsection{Training and Evaluation Splits}

We separate policy training from evaluation along task, network-size, and topology dimensions. The training environments contain small graphs with network sizes $N\in\{10,50\}$ and use Random and Scale-Free topologies. Networks with $N\in\{100,500,1000\}$ are reserved for size generalization, while Semantic networks are held out to evaluate topology generalization.

Within the training environments, we additionally reserve $10\%$ of task identifiers as an in-distribution test set. The holdout is deterministic and based on a hash of the task identifier, ensuring that the same task is consistently assigned to the same split across network configurations and repeated experiments. Held-out task identifiers are excluded from all SFT examples and from online OPSD and GRPO training.

This produces three evaluation regimes: in-distribution generalization to unseen tasks, size extrapolation to larger graphs, and topology transfer to a graph family not used during training. All trained policies are evaluated by executing them live in the corresponding held-out environments.

\subsubsection{Shared Local Policy}

A collaboration trajectory contains a sequence of visits to potentially different agents. At each visit, the acting agent receives a system instruction together with its current local observation. Depending on the benchmark configuration, this observation contains the current request, locally owned tools, immediate neighbors, available intermediate results, and the interaction history visible to that agent.

The model emits a structured decision specifying the actions to execute, such as returning a local result or forwarding the request to selected neighbors. For offline SFT data, we train on the parsed decision accepted by the environment rather than the teacher model's unrestricted natural-language response. Accepted decisions are canonicalized into a consistent JSON representation. This removes stylistic variation among data-generating models and focuses learning on the action that changes the environment.

The same structured action representation is used during online OPSD and GRPO. Consequently, differences among the three training methods arise from how training states and learning signals are obtained, rather than from differences in the underlying agent interface.

\subsubsection{Training Reward}
\label{sec:training-reward}

We use the benchmark evaluator to assign an outcome score to each complete collaboration trajectory. The reward combines task correctness, collaboration quality, and efficiency. Let $S$, $C$, and $E$ denote these components normalized to $[0,1]$.

For offline trajectory selection, we use

\begin{equation}
r_{\mathrm{offline}}
=
\frac{S+C+3E}{5}.
\end{equation}

The larger weight on efficiency discourages flooding. A policy that forwards every request broadly can often obtain the required resource and therefore achieve reasonable correctness and collaboration scores, despite exhibiting poor decentralized routing. Weighting efficiency more strongly favors trajectories that discover the required resources selectively.

For online policy optimization, we use a success-gated variant,

\begin{equation}
r_{\mathrm{online}}
=
S
\frac{C+3E}{4}.
\end{equation}

Gating by correctness prevents the policy from obtaining high reward merely by terminating early or minimizing communication without solving the task.

The training correctness signal uses exact per-tool output accuracy, which provides dense partial credit when only a subset of the required tools is successfully obtained. This dense signal is used for policy training only. Headline evaluation continues to use the task success, collaboration, and efficiency metrics described in Section~\ref{sec:metrics}.

For a fully local task, no remote collaboration is required. To avoid granting a constant collaboration reward in such cases, we replace the collaboration component with the task correctness signal during training.

\subsubsection{Supervised Fine-Tuning}
\label{sec:sft-details}

SFT provides the initial learned collaboration policy. We construct its dataset from previously collected trajectories on the training environments.

\paragraph{Trajectory filtering.}
Each completed episode is assigned the offline reward defined above. We retain episodes whose reward is at least $0.5$. We exclude trajectories generated by the Broadcast policy because indiscriminate forwarding is an undesirable behavior that we do not want to imitate. We also exclude Oracle Route and Oracle Route All trajectories because their prompts contain routing information that is unavailable under the normal decentralized observation model.

Only episodes belonging to the training split are used. The held-out in-distribution task subset is excluded.

\paragraph{Node-level examples.}
For every accepted decision in a retained episode, we create one supervised example. The input contains the system instruction and local observation seen at that node. The completion is the canonicalized structured decision executed by the environment.

Thus, an episode visiting several agents contributes several training examples. All decisions from the same episode inherit its trajectory-level reward for filtering purposes.

Invalid outputs and attempts that are rejected by the environment are not used as positive demonstrations.

\paragraph{Objective.}
Let $o$ denote the local observation and $a=(a_1,\ldots,a_L)$ the serialized structured decision. SFT minimizes the standard autoregressive negative log-likelihood of the decision tokens:

\begin{equation}
\mathcal{L}_{\mathrm{SFT}}
=
-\sum_{\ell=1}^{L}
\log \pi_\theta(a_\ell \mid o,a_{<\ell}).
\end{equation}

The loss is applied only to the assistant completion containing the decision. Tokens belonging to the system instruction and local observation are masked from the training objective.

\paragraph{Configuration.}
We initialize SFT from Qwen3.5-4B-Instruct. Unless otherwise specified, we train for two epochs with learning rate $1\times10^{-5}$, per-device batch size $8$, gradient accumulation over $4$ batches, and maximum sequence length $8192$. Training uses bfloat16 precision and gradient checkpointing. The training examples are shuffled before tokenization.

The default SFT experiment performs full-model fine-tuning. The implementation additionally supports LoRA with rank $32$, scaling parameter $64$, dropout $0.05$, and adapters on the query, key, value, and output projection modules.

\subsubsection{On-Policy Self-Distillation}
\label{sec:OPSD-details}

SFT learns only from states present in the fixed offline trajectory corpus. Once deployed, however, the student may make different routing decisions and reach local states that are rare or absent from this dataset. OPSD addresses this distribution mismatch by collecting supervision under the student's current policy.

\paragraph{Initialization.}
We initialize the student policy $\pi_\theta$ from the SFT checkpoint. OPSD uses the same model as a self-teacher, augmented with oracle routing information, to provide supervision. The teacher does not control the trajectory.

In particular, when the student reaches agent $A_i$, the self-teacher receives the same local observation $o_{i,t}$ together with oracle routing information for the required resources, including their owners and shortest routes. The student receives only the standard local observation and does not have access to this privileged routing information.

The oracle information provides a stronger supervision signal while remaining hidden from the student at inference time.

\paragraph{On-policy state collection.}
At OPSD iteration $k$, the current student policy $\pi_{\theta_k}$ is deployed in the training environments. The student, rather than the teacher, chooses the actions that determine how the trajectory evolves:

\begin{equation}
a_{i,t}
\sim
\pi_{\theta_k}(\cdot \mid o_{i,t}).
\end{equation}

The resulting observations therefore follow the state distribution induced by the current student policy.

For every visited observation $o_{i,t}$, we separately query the teacher using the same local context. The teacher produces a target decision $a^T_{i,t}$. The resulting OPSD example is

\begin{equation}
(o_{i,t},a^T_{i,t}).
\end{equation}

The teacher action is used as a training target but is not used to advance the environment during data collection. Consequently, subsequent states remain genuinely on-policy with respect to the student.

\paragraph{Distillation objective.}
The student is optimized to match the teacher on these student-induced states. With token-level teacher supervision, the OPSD objective is

\begin{equation}
\mathcal{L}_{\mathrm{OPSD}}
=
-\mathbb{E}_{o\sim d_{\pi_\theta}}
\left[
\sum_{\ell=1}^{L}
\log
\pi_\theta
\left(
a^T_\ell
\mid
o,a^T_{<\ell}
\right)
\right],
\end{equation}

where $d_{\pi_\theta}$ denotes the observation distribution induced by the current student policy.

If teacher token probabilities are available, the same procedure can instead use a token-level KL distillation loss between the teacher and student distributions. In our structured-action setting, the essential distinction from SFT is not the particular token-level divergence, but that supervision is generated on observations sampled from the current student policy.

\paragraph{Iterative training.}
OPSD alternates between rollout and optimization. At each iteration, we:

\begin{enumerate}
    \item deploy the current student on the training networks
    \item collect the local observations encountered by the student
    \item query the frozen teacher on each collected observation
    \item canonicalize valid teacher decisions into the benchmark action format
    \item update the student using the newly collected teacher targets
\end{enumerate}

The updated policy is then used to generate the next iteration of trajectories. Thus, as the student's routing behavior changes, the distribution of states receiving teacher supervision changes with it.

\paragraph{Why on-policy distillation.}
This distinction is particularly important in decentralized routing because early routing mistakes change all subsequent observations. An offline dataset may contain strong decisions along successful routes but provide little supervision for recovery states produced by the student itself. OPSD exposes the teacher to precisely those states, including states reached after suboptimal forwarding choices, and teaches the student how a stronger policy would act from the same locally observable situation.

OPSD nevertheless remains a distillation method rather than reward optimization. The target at each state is the teacher's local decision. Environment reward can be used to monitor or filter rollouts, but it is not the direct optimization signal.

\paragraph{Configuration.}
We initialize OPSD from the same SFT checkpoint used to initialize GRPO. Student rollouts use only the training networks and training task identifiers. The self-teacher is frozen throughout training and is queried deterministically unless otherwise specified. The student is updated after each collection round using only the teacher completion tokens.

For a controlled comparison with GRPO, we use the same local observation format, training environments, maximum context length, and structured decision representation. OPSD never observes held-out Semantic networks, larger test networks, or held-out task identifiers during training.

\subsubsection{Group Relative Policy Optimization}
\label{sec:grpo-details}

GRPO directly optimizes collaboration outcomes through online environment interaction. As with OPSD, we initialize the policy from the SFT checkpoint. Unlike OPSD, no teacher action is provided. Instead, the policy learns from relative rewards among multiple trajectories sampled for the same task.

\paragraph{Grouped rollouts.}
For each training task, the current policy generates a group of $G$ complete decentralized trajectories. We use $G=8$ by default. Sampling uses temperature $1.0$.

Each rollout executes the normal benchmark environment. All intermediate routing decisions are therefore made by the current policy using the same local observations available at evaluation time.

For trajectory $g$, the environment produces reward $r_g$. We normalize rewards within each rollout group to obtain a group-relative advantage. Conceptually,

\begin{equation}
A_g
=
\frac{r_g-\mu_r}
{\sigma_r+\epsilon},
\end{equation}

where $\mu_r$ and $\sigma_r$ are the mean and standard deviation of rewards among trajectories generated for the same task.

Every local decision belonging to trajectory $g$ receives the corresponding trajectory-level advantage $A_g$. This provides credit to decisions participating in relatively successful collaboration trajectories and negative signal to decisions participating in poorer trajectories.

\paragraph{Policy objective.}
For each collected local decision, we compute the mean log probability of the completion tokens under the current policy. The GRPO update increases the likelihood of actions from positive-advantage trajectories and decreases the likelihood of actions from negative-advantage trajectories.

We additionally regularize the policy toward the SFT initialization:

\begin{equation}
\mathcal{L}_{\mathrm{GRPO}}
=
-\mathbb{E}
\left[
A\log\pi_\theta(a\mid o)
\right]
+
\beta
\mathbb{E}
\left[
\log\pi_\theta(a\mid o)
-
\log\pi_{\mathrm{ref}}(a\mid o)
\right].
\end{equation}

The reference policy $\pi_{\mathrm{ref}}$ is the frozen SFT model.

\paragraph{Round-based optimization.}
Because a benchmark episode contains multi-agent, multi-turn interaction with an external environment, training uses an explicit rollout-and-update loop rather than an in-process single-turn GRPO trainer.

Each training round performs four stages:

\begin{enumerate}
    \item create a servable snapshot of the current policy
    \item execute grouped rollouts in the training environments
    \item compute environment rewards and group-relative advantages
    \item update the policy on the collected local decisions
\end{enumerate}

The updated policy is then used for the next rollout round.

\paragraph{Parameter-efficient update.}
GRPO applies LoRA to the SFT model. We use rank $32$, scaling parameter $64$, and adapters on the query, key, value, and output projections. The underlying SFT parameters remain frozen. This design also permits the frozen model with the adapter disabled to serve as the reference policy for KL regularization without maintaining a second full model in memory.

\paragraph{Configuration.}
The default configuration uses $20$ GRPO rounds, group size $8$, sampling temperature $1.0$, learning rate $1\times10^{-6}$, and KL coefficient $\beta=0.02$. Updates use micro-batch size $1$ and a maximum context length of $4096$. Longer contexts are left-truncated so that the structured completion at the end of the sequence is always retained.

We use AdamW for optimization and clip the norm of trainable gradients to $1.0$. The online reward uses the success-gated formulation described in Section~\ref{sec:training-reward} to reduce efficiency reward hacking.

\subsubsection{Comparison of Training Signals}

The three methods differ primarily in where their training states and supervision originate.

SFT uses a fixed offline state distribution generated by previously collected high-quality trajectories and imitates their accepted actions. OPSD generates states using the current student policy but obtains the target action from a self-teacher with oracle routing information. GRPO also generates states using the current policy, but replaces teacher supervision with trajectory-level environment reward.

Thus, the progression from SFT to OPSD to GRPO separates three increasingly interactive forms of policy improvement: offline imitation, on-policy imitation, and on-policy reward optimization.

\section{Additional Experimental Details}
\label{sec:additional-experiments}

This appendix provides the detailed evaluation protocol, baseline definitions, ablations, and training objective used in the decentralized collaboration experiments.

\subsection{Entry-Agent Evaluation by Resource Distance}
\label{sec:entry-distance}

For each task, we evaluate every agent as the entry agent and group runs by the distance to the required resources. For a task $q$ with required tool set $\mathcal{T}(q)$, we define

\begin{equation}
d(q,A_0)=\max_{t\in\mathcal{T}(q)} d_{\mathcal{G}}\left(A_0,o(t)\right),
\end{equation}

where $o(t)$ is the owner of tool $t$. The maximum distance captures the farthest resource that must be reached to complete the task. We report four primary buckets:
\begin{itemize}
    \item \textbf{Local}: $d(q,A_0)=0$, so all required tools are owned by the entry agent.
    \item \textbf{1-hop}: $d(q,A_0)=1$.
    \item \textbf{2-hop}: $d(q,A_0)=2$.
    \item \textbf{3+-hop}: $d(q,A_0)\geq 3$.
\end{itemize}

The headline benchmark uses connected graphs, so every required tool is reachable from every entry agent. If disconnected graphs are introduced as a robustness condition, unreachable cases are reported separately and excluded from headline task success because the task is not solvable from that entry point.

Evaluating every entry agent makes a separate random-reachable bucket unnecessary. If exhaustive evaluation becomes too expensive at larger scales, the same analysis can be approximated by stratified sampling within these distance buckets.

\subsection{Request Semantics and Within-Task Cache}
\label{sec:request-cache}

\paragraph{Request semantics.}
The headline tool benchmark uses result requests. A queried owner executes its local tool and returns only the output. This matches the benchmark action space in which \texttt{Return} exposes a tool result rather than the private tool implementation.

We additionally support a content-transfer ablation. Under a \texttt{content} request, an owner may return transferable resource content that the requester can mount and execute locally. Under a \texttt{result} request, execution remains delegated to the owner. The request specifies what the caller prefers, while the queried agent may still reject or forward according to its policy. This ablation separates open resource sharing from service-style delegation.

\paragraph{Entry-agent cache.}
A task may require several resources across multiple pipeline stages. We therefore distinguish a within-task cache from the cross-task experience memory used by \method. The within-task cache may store discovered resource locations, previous responses, transferred resource bodies when transfer is enabled, and intermediate execution results.

We compare three cache policies. Under \texttt{agent}, the entry agent decides what to retain. Under \texttt{lru}, eviction follows a fixed least-recently-used policy. Under \texttt{off}, no reusable cache is maintained. This ablation measures whether explicit within-task memory reduces repeated queries and redundant execution.

\subsection{Routing and Coordination Ablation}
\label{sec:routing-coordination}

We compare two coordination regimes while preserving neighbor-only communication.

\paragraph{Entry-controlled routing.}
The entry agent remains the decision maker for the route. Intermediate agents expose only the local information needed for the entry agent to choose the next hop and then relay the request along that selected edge. Intermediate agents do not independently choose where to forward. This condition tests routing under a single task-level controller without granting global network access.

\paragraph{Recursive routing.}
Each receiving agent independently decides whether to answer, reject, or forward the request to one or more of its own neighbors. The entry agent may therefore receive a result without observing every intermediate routing decision.

The first condition isolates the routing ability of the entry agent. The second evaluates genuinely distributed coordination among multiple local decision makers. Recursive routing is the default condition in the main benchmark.

\subsection{Memory and Network Dynamics}
\label{sec:memory-dynamics}

Persistent routing information is useful only when it remains sufficiently aligned with the network. We therefore vary both cross-task memory and network dynamics.

\paragraph{Cross-task memory.}
Under \texttt{none}, each task begins with an empty cross-task routing state. Under \texttt{persistent}, agents retain routing notes, resource-location estimates, previous outcomes, and neighbor-performance statistics across tasks. This memory is distinct from the within-task cache in Section~\ref{sec:request-cache}.

Persistent-memory evaluation is an online adaptation setting. All methods that retain cross-task state should observe the same task order. To reduce sensitivity to a particular order, results should be averaged over multiple task-order permutations.

\paragraph{Static graph.}
Resource ownership and communication edges remain fixed throughout evaluation. Persistent memory can therefore accumulate stable routing information.

\paragraph{Dynamic graph between tasks.}
Between consecutive tasks, a controlled fraction of communication edges is perturbed and a controlled fraction of resources migrates to new owners. This setting tests whether agents learn transferable routing patterns or only memorize fixed locations.

\paragraph{Dynamic graph within tasks.}
As a stronger stress test, edges or resource owners may change between communication rounds while a task is being solved. Each update preserves graph connectivity and exactly one owner for every active tool. Changes occur only at round boundaries so that the environment remains well defined. This condition tests whether agents can detect stale routing information and recover online.

\subsection{\method Warmup and Online Adaptation}
\label{sec:method-details}

\method separates pre-evaluation topology acquisition from online routing adaptation.

\paragraph{Warmup data.}
Warmup uses auxiliary tasks that are disjoint from the test tasks. Agents may exchange compact information about advertised capabilities, local routing outcomes, and local network structure. Private tool implementations and secret execution keys remain hidden.

Each agent starts from a local state $W_i^{(0)}$. One warmup layer performs

\begin{equation}
\small
W_i^{(\ell+1)}=\Phi\left(W_i^{(\ell)},\operatorname{AGG}\left\{\Psi\left(W_i^{(\ell)},W_j^{(\ell)}\right):A_j\in\mathcal{N}_i\right\}\right).
\end{equation}

After $L$ rounds, $W_i=W_i^{(L)}$ is fixed for evaluation. The update is GNN-style in the sense that information propagates through repeated local aggregation. It does not require a globally materialized graph representation at any agent.

\paragraph{Online routing memory.}
During evaluation, agent $A_i$ maintains local experience memory $M_{i,t}$. For a request involving resource $k$ and candidate neighbor $A_j$, the LLM computes

\begin{equation}
s_i(k,A_j,t)=f_{\mathrm{LLM}}\left(k,A_j,W_i,M_{i,t},h_{i,t}\right).
\end{equation}

We use the notation $s_i$ rather than a Q-value because this score is generated from contextual routing evidence and is not constrained by a Bellman update. After observing the local routing outcome, the agent updates its memory

\begin{equation}
\small
M_{i,t+1}=\operatorname{Update}_{\mathrm{LLM}}\left(M_{i,t},k,A_j,r_i,o_{i,t+1}\right).
\end{equation}

This distinction separates \method from the numerical Q-Routing baseline. Numerical Q-Routing adapts a tabular scalar for each observed resource-neighbor pair. \method uses the LLM to generalize from local routing context and stored experience across semantically related requests.

\subsection{Training Objective for Learned Policies}
\label{sec:training-reward}

For policies trained with reinforcement learning, we use a scalar reward that combines partial task correctness with interaction cost. Let $N$ be the total number of node visits, $U$ the number of unique visited nodes, $C$ the number of constraint violations, and $P$ the number of repeated node visits. Let $\mathrm{ToolAcc}$ denote mean per-tool output accuracy. The training reward is

\begin{equation}
\small
R=\mathrm{ToolAcc}-\lambda_N N-\lambda_U U-\lambda_C C-\lambda_P P.
\end{equation}

The default coefficients are

\begin{equation}
\small
\lambda_N=0.01,\qquad \lambda_U=0,\qquad \lambda_C=0.5,\qquad \lambda_P=0.05.
\end{equation}

The dense correctness term is used only as a training signal. Headline evaluation continues to use binary end-to-end task success. The coefficients are configurable, and individual cost metrics are reported separately rather than being replaced by this scalar reward.

\section{Additional Results}
\label{app:results}

We present the full results across different LLMs in Table~\ref{tab:main_full}.

\begin{table*}[t]
\centering
\caption{Performance comparison across agent designs and base LLMs.}
\label{tab:main_full}
\resizebox{\linewidth}{!}{
\begin{tabular}{llccc}
\toprule
\textbf{Agent} & \textbf{Base LLM} & \textbf{Success ($\uparrow$)} & \textbf{Collaboration Quality ($\uparrow$)} & \textbf{Efficiency ($\uparrow$)} \\
\midrule
% \midrule
% \multicolumn{5}{c}{\textit{Centralized}} \\
% \midrule
\rowcolor{Gray} Global Registry       & NA & \multicolumn{3}{c}{Out of context} \\
% \rowcolor{Gray} Global Registry       & claude-sonnet-4-6 & \multicolumn{3}{c}{Out of context} \\
% \rowcolor{Gray} Global Registry + RAG & claude-sonnet-4-6 & 82.6 & N/A & N/A \\
% \midrule
% \multicolumn{5}{c}{\textit{Decentralized}} \\
% \midrule
\midrule
\multicolumn{5}{c}{\textit{Static Agents}} \\
\midrule
% Reflex & claude-haiku-4-5   & 5.0 & 32.1 & 69.4 \\
% Reflex & claude-sonnet-4-6  & 6.4 & 34.4 & 59.9 \\
% Reflex & claude-sonnet-5    & 2.5 & 17.6 & 91.2 \\
% Reflex & claude-opus-4-8    & 7.3 & 31.7 & 71.2 \\
% Reflex & openai.gpt-5.4     & 18.9 & 44.6 & 46.9 \\
% Reflex & openai.gpt-5.5     & 3.0 & 8.7 & 92.8 \\
% Reflex & openai.gpt-5.6-sol & 95.3 & 95.3 & 41.9 \\
Reflex & Gemma 12B         & 5.0 & 32.1 & 69.4 \\
Reflex & Gemma 4 26B A4B   & 6.4 & 34.4 & 59.9 \\
Reflex & Muse Glimmer 30B  & 2.5 & 17.6 & 91.2 \\
Reflex & Qwen3.5-0.8B       & 1.5 & 14.9 & 22.6 \\
Reflex & Qwen3.5 35B A3B    & 3.3 & 21.1 & 25.7 \\
Reflex & QWen3.8 27B        & 5.2 & 27.5 & 37.5 \\
\midrule
% Reflex + History & claude-haiku-4-5 & 3.7 & 7.1 & 97.2 \\
% Reflex + History & claude-sonnet-4-6 & 6.2 & 28.6 & 72.3 \\
% Reflex + History & claude-sonnet-5 & 2.9 & 13.4 & 93.6 \\
% Reflex + History & claude-opus-4-8 & 9.3 & 30.9 & 82.2 \\
% Reflex + History & openai.gpt-5.4 & 2.6 & 11.8 & 87.8 \\
% Reflex + History & openai.gpt-5.5 & 2.6 & 3.7 & 98.6 \\
% Reflex + History & openai.gpt-5.6-luna & 40.3 & 58.8 & 50.2 \\
% Reflex + History & openai.gpt-5.6-terra & 40.9 & 71.8 & 47.4 \\
% Reflex + History & openai.gpt-5.6-sol & 57.3 & 16.0 & 100.0 \\
Reflex + History & Gemma 12B         & 3.7 & 7.1 & 97.2 \\
Reflex + History & Gemma 4 26B A4B   & 6.2 & 28.6 & 72.3 \\
Reflex + History & Muse Glimmer 30B  & 2.9 & 13.4 & 93.6 \\
Reflex + History & Qwen3.5-0.8B       & 2.1 & 16.2 & 25.1 \\
Reflex + History & Qwen3.5 35B A3B    & 3.6 & 24.0 & 31.5 \\
Reflex + History & QWen3.8 27B        & 5.0 & 30.3 & 36.1 \\
\midrule
% Broadcast & claude-haiku-4-5   & 56.2 & 76.1 & 38.8 \\
% Broadcast & claude-sonnet-4-6  & 71.3 & 75.3 & 36.4 \\
% Broadcast & claude-sonnet-5    & 72.8 & 80.1 & 36.7 \\
% Broadcast & claude-opus-4-8    & 83.1 & 77.9 & 35.4 \\
% Broadcast & openai.gpt-5.4     & 68.5 & 86.2 & 29.8 \\
% Broadcast & openai.gpt-5.5     & 75.3 & 80.3 & 33.6 \\
% Broadcast & openai.gpt-5.6-sol & 100.0 & 93.1 & 31.6 \\
Broadcast & Gemma 12B          & 56.2 & 76.1 & 38.8 \\
Broadcast & Gemma 4 26B A4B    & 71.3 & 75.3 & 36.4 \\
Broadcast & Muse Glimmer 30B   & 72.8 & 80.1 & 36.7 \\
Broadcast & Qwen3.5-0.8B       & 15.7 & 66.8 & 45.1 \\
Broadcast & Qwen3.5 35B A3B    & 20.6 & 70.9 & 46.3 \\
Broadcast & QWen3.8 27B        & 40.6 & 68.7 & 45.7 \\
\midrule
% Oracle Route (Entry Only) & claude-haiku-4-5 & 33.3 & 62.6 & 65.9 \\
% Oracle Route (Entry Only) & claude-sonnet-4-6 & 33.5 & 62.1 & 69.5 \\
% Oracle Route (Entry Only) & claude-opus-4-8 & 33.6 & 63.9 & 68.7 \\
Oracle Route (Entry Only) & Gemma 12B          & 33.3 & 62.6 & 65.9 \\
Oracle Route (Entry Only) & Gemma 4 26B A4B    & 33.5 & 62.1 & 69.5 \\
Oracle Route (Entry Only) & Muse Glimmer 30B   & 33.6 & 63.9 & 68.7 \\
% Oracle Route (All Agents) & claude-haiku-4-5 & 13.1 & 40.5 & 68.5 \\
% Oracle Route (All Agents) & claude-sonnet-4-6 & 27.7 & 59.3 & 67.2 \\
% Oracle Route (All Agents) & claude-sonnet-5 & 5.1 & 44.3 & 70.1 \\
% Oracle Route (All Agents) & claude-opus-4-8 & 30.1 & 41.1 & 82.9 \\
Oracle Route (All Agents) & Gemma 12B        & 13.1 & 40.5 & 68.5 \\
Oracle Route (All Agents) & Gemma 4 26B A4B  & 27.7 & 59.3 & 67.2 \\
Oracle Route (All Agents) & Muse Glimmer 30B & 5.1 & 44.3 & 70.1 \\
\midrule
\multicolumn{5}{c}{\textit{Self-Improved Agents}} \\
\midrule
Numeric Q-Routing, Learn w/ All & Gemma 12B        & 5.0 & 22.0 & 80.1 \\
Numeric Q-Routing, Learn w/ All & Gemma 4 26B A4B  & 3.9 & 21.2 & 80.8 \\
Numeric Q-Routing, Learn w/ All & Muse Glimmer 30B & 1.7 & 2.5 & 99.3 \\
% Numeric Q-Routing, Learn w/ All & claude-haiku-4-5 & 5.0 & 22.0 & 80.1 \\
% Numeric Q-Routing, Learn w/ All & claude-sonnet-4-6 & 3.9 & 21.2 & 80.8 \\
% Numeric Q-Routing, Learn w/ All & claude-sonnet-5 & 1.7 & 2.5 & 99.3 \\
% Numeric Q-Routing, Learn w/ All & claude-opus-4-8 & 8.4 & 18.4 & 86.2 \\
% Numeric Q-Routing, Learn w/ All & openai.gpt-5.4 & 1.2 & 3.8 & 93.1 \\
% Numeric Q-Routing, Learn w/ All & openai.gpt-5.5 & 1.6 & 5.6 & 93.5 \\
% Numeric Q-Routing, Learn w/ All & openai.gpt-5.6-sol & 91.1 & 94.7 & 32.0 \\
\midrule
Success-Trajectory Memory & Gemma 12B        &  1.1 & 22.3 & 62.5 \\
Success-Trajectory Memory & Gemma 4 26B A4B  &  1.1 & 31.0 & 82.3 \\
Success-Trajectory Memory & Muse Glimmer 30B &  1.6 & 26.3 & 57.8 \\
% Success-Trajectory Memory & claude-haiku-4-5   &  1.1 & 22.3 & 62.5 \\
% Success-Trajectory Memory & claude-sonnet-4-6  &  1.1 & 31.0 & 82.3 \\
% Success-Trajectory Memory & claude-sonnet-5    &  1.6 & 26.3 & 57.8 \\
% Success-Trajectory Memory & claude-opus-4-8    &  2.8 & 11.6 & 100.0 \\
% Success-Trajectory Memory & openai.gpt-5.4     &  2.0 & 25.1 & 39.5 \\
% Success-Trajectory Memory & openai.gpt-5.5     &  2.1 & 30.3 & 44.9 \\
% Success-Trajectory Memory & openai.gpt-5.6-sol & 95.7 & 40.6 & 78.3 \\

\midrule

Location Learning & Gemma 12B         & 3.6 & 36.1 & 56.8 \\
Location Learning & Gemma 4 26B A4B   & 3.8 & 46.7 & 53.3 \\
Location Learning & Muse Glimmer 30B  & 5.3 & 50.6 & 71.5 \\
% Location Learning & claude-haiku-4-5   & 3.6 & 36.1 & 56.8 \\
% Location Learning & claude-sonnet-4-6  & 3.8 & 46.7 & 53.3 \\
% Location Learning & claude-sonnet-5    & 5.3 & 50.6 & 71.5 \\
% Location Learning & claude-opus-4-8    & 8.5 & 20.3 & 86.9 \\
% Location Learning & openai.gpt-5.4     & 4.1 & 38.7 & 66.2 \\
% Location Learning & openai.gpt-5.5     & 4.6 & 53.5 & 62.7 \\
% Location Learning & openai.gpt-5.6-sol & 97.3 & 96.3 & 45.1 \\

\midrule
LLM Q-Learning (Our) & Gemma 12B         & 17.5 & 33.8 & 47.5 \\
LLM Q-Learning (Our) & Gemma 4 26B A4B   & 19.1 & 35.9 & 63.1 \\
LLM Q-Learning (Our) & Muse Glimmer 30B  & 20.6 & 37.7 & 56.8 \\

% LLM Q-Learning (Our) & claude-haiku-4-5   & 17.5 & 33.8 & 47.5 \\
% LLM Q-Learning (Our) & claude-sonnet-4-6  & 19.1 & 35.9 & 63.1 \\
% LLM Q-Learning (Our) & claude-sonnet-5    & 20.6 & 37.7 & 56.8 \\
% LLM Q-Learning (Our) & claude-opus-4-8    & 46.0 & 60.2 & 73.1 \\
% LLM Q-Learning (Our) & openai.gpt-5.4     & 16.6 & 36.2 & 53.1 \\
% LLM Q-Learning (Our) & openai.gpt-5.5     & 17.1 & 39.5 & 50.4 \\
% LLM Q-Learning (Our) & openai.gpt-5.6-sol & 96.8 & 55.3 & 58.3 \\
% LLM Q-Learning (Our) & Qwen3.5-0.8B       &  6.6 & 48.6 & 58.3 \\
% LLM Q-Learning (Our) & Qwen3.5 35B A3B    &  7.3 & 55.1 & 61.2 \\
% LLM Q-Learning (Our) & QWen3.8 27B        & 11.2 & 57.5 & 61.7 \\

\midrule

% Warmup 2 Hop + LLM Q-Learning (Our) & claude-sonnet-4-6 & \textbf{76.8} & \textbf{78.3} & \textbf{83.8} \\
Warmup 2 Hop + LLM Q-Learning (Our) & Gemma 12B        & 28.1 & 45.5 & 59.8 \\
Warmup 2 Hop + LLM Q-Learning (Our) & Gemma 4 26B A4B  & 30.6 & 49.1 & 40.5 \\
Warmup 2 Hop + LLM Q-Learning (Our) & Muse Glimmer 30B & 38.9 & 57.2 & 66.3 \\
Warmup 2 Hop + LLM Q-Learning (Our) & Qwen3.5-0.8B      & 35.1 & 49.7 & 56.7 \\
Warmup 2 Hop + LLM Q-Learning (Our) & Qwen3.5 35B A3B   & 45.4 & 55.3 & 64.9 \\
Warmup 2 Hop + LLM Q-Learning (Our) & QWen3.8 27B       & 53.7 & 58.1 & 71.5 \\

\midrule
\textbf{Trained Agents} \\

SFT  & Qwen3.5-0.8B & 0.4 & 49.5 & 31.8 \\
SFT  &   Qwen3.5-2B & 0.6 & 50.2 & 33.5 \\
SFT  &   Qwen3.5-4B & 0.8 & 48.0 & 36.3 \\

GRPO & Qwen3.5-0.8B & 0.2 & 38.6 & 51.5 \\
GRPO &   Qwen3.5-2B & 0.2 & 45.7 & 57.2 \\
GRPO &   Qwen3.5-4B & 0.5 & 46.8 & 56.6 \\

OPSD & Qwen3.5-0.8B & 3.3 & 36.7 & 60.2 \\
OPSD &   Qwen3.5-2B & 3.6 & 44.6 & 53.7 \\
OPSD &   Qwen3.5-4B & 5.7 & 46.9 & 56.4 \\

Reflex & Qwen3.5-0.8B       & 1.5 & 14.9 & 22.6 \\
Reflex & Qwen3.5 35B A3B    & 3.3 & 21.1 & 25.7 \\
Reflex & QWen3.8 27B        & 5.2 & 27.5 & 37.5 \\

\bottomrule
\end{tabular}
}
\end{table*}

\end{document}